\documentclass[prb,showpacs,twocolumn,floatfix]{revtex4} 
\usepackage{multirow}
\usepackage{amsmath}
\usepackage{graphicx}
\usepackage{comment}
\usepackage{float}
\usepackage{amssymb}
\usepackage{times}
\usepackage[colorlinks,hyperindex]{hyperref}
\DeclareGraphicsExtensions{.eps,.ps}
\hypersetup
	{
		colorlinks,%
		citecolor=black,%
		linkcolor=blue,%
		urlcolor=black,%
	}

\begin{document}
\title{Dynamic scaling at classical phase transitions approached through non-equilibrium quenching}

\author{Cheng-Wei Liu}

\author{Anatoli Polkovnikov}

\author{Anders W. Sandvik}

\affiliation{Department of Physics, Boston University, 590 Commonwealth Avenue, Boston, Massachusetts 02215}

\date{\today}
 
\begin{abstract}
We use Monte Carlo simulations to demonstrate generic scaling aspects of classical phase transitions approached through a quench (or annealing) 
protocol where the temperature changes as a function of time with velocity $v$. Using a generalized Kibble-Zurek ansatz, we demonstrate dynamic 
scaling for different types of stochastic dynamics (Metropolis, Swendsen-Wang, and Wolff) on Ising models in two and higher dimensions. We show that 
there are dual scaling functions governing the dynamic scaling, which together describe the scaling behavior in the entire velocity range $v \in [0, \infty)$.
These functions have asymptotics corresponding to the adiabatic and diabatic limit, and close to these limits they are perturbative in $v$ and $1/v$, 
respectively. Away from their perturbative domains, both functions cross over into the same universal power-law 
scaling form governed by the static and dynamic critical exponents (as well as an exponent characterizing the quench protocol). As a by-product of 
the scaling studies, we obtain high-precision estimates of the dynamic exponent $z$ for the two-dimensional Ising model subject to the three variants 
of Monte Carlo dynamics; for single-spin Metropolis updates $z_M=2.1767(5)$, for Swendsen-Wang multi-cluster updates $z_{\rm SW}=0.297(3)$, and for Wolff 
single-cluster updates $z_{W}=0.30(2)$. For Wolff dynamics, we find an interesting behavior with a non-analytic break-down of the quasi-adiabatic 
and diabatic scaling, instead of the generic smooth cross-over described by a power law. We interpret this disconnect between the two scaling 
regimes as a dynamic phase transition of the Wolff algorithm, caused by an effective sudden loss of ergodicity at high velocity.
\end{abstract}

\pacs{64.60.De, 64.60.F-, 64.60.Ht, 05.70.Ln} 
\maketitle

\section{Introduction}
\label{sec:intro}

Phase transitions and critical phenomena have formed a dominant theme in statistical physics for a long time and new aspects are still subject 
to active research. This is not only because of the importance and elegance of the fundamental aspect of many-body systems in the original setting 
of condensed-matter physics, but also thanks to diverse applications to various complex systems in other areas of physics, as well as in chemistry, 
biology, and even in economy and social sciences. Any system with collective behavior resulting from a large number of interacting particles 
(or ``agents'') can be described by methods of statistical physics, and phase transitions often are important features of such systems.

A fundamental aspect of phase transitions is the scale invariance emerging upon approaching a critical point, which leads to universal scaling 
behavior independent of microscopic characteristics. The theoretical understanding of universality in equilibrium statistical mechanics is well established 
in terms of the renormalization group (RG).\cite{RevModPhys.47.773} Attempts have also been made to generalize this formalism as well as general scaling
hypotheses to non-equilibrium phase transitions and dynamic critical scaling,
\cite{RevModPhys.49.435,J.Phys.A.9.1387,Nature.317.505,Janssen89,PRB.71.132402,PhysRevB.72.161201,PhysRevLett.95.105701,PhysRevLett.95.245701,Gong10,PhysRevB.84.224303, EPL.84.67008, PhysRevB.86.064304,PhysRevLett.109.015701, PhysRevLett.104.160404, chandran_2012, PhysRevB.85.100505, qaqmc, Dziarmaga2010,RevModPhys.83.863}
but the understanding here is much less complete. Since many important systems are far 
from equilibrium, deeper understanding of criticality and scaling behavior under such conditions is called for. 

In this paper we report progress in characterizing dynamical critical scaling at classical (thermal) phase transitions. We discuss a scaling hypothesis for a very 
general class of quench (or annealing) protocols in which a function with a single dynamic exponent (along with the standard equilibrium exponents) describes the 
changes from adiabatic to diabatic evolution. To test the scaling forms we study phase transitions in classical Ising models, using Monte Carlo (MC) simulations 
with both single-spin and cluster updates.

\subsection{Kibble-Zurek Mechanism}
\label{subsec:kz}

Our approach is based on extensions of the Kibble-Zurek (KZ) arguments,\cite{J.Phys.A.9.1387,Nature.317.505} which originally focused on quantitatively 
relating defect formation (e.g., the typical defect size and the density of defects) to the rate of change (the quench velocity) of a parameter of the system 
(such as the temperature, external fields, etc.). The KZ mechanism and extensions of it have successfully been used to describe out-of-equilibrium physics 
at both classical \cite{J.Phys.A.9.1387,Nature.317.505,PhysRevB.86.064304} and quantum phase transitions 
\cite{EPL.84.67008, PhysRevB.72.161201, PhysRevLett.95.105701, PhysRevLett.95.245701, PhysRevB.84.224303, PhysRevLett.109.015701, chandran_2012, PhysRevB.85.100505, qaqmc} 
(for a general review, see, e.g., Refs.~\onlinecite{RevModPhys.83.863, Dziarmaga2010}). 

We consider a system with critical temperature $T_c$.
When this system is quenched to the neighborhood of $T_c$ by starting from some initial temperature $T_i>T_c$ and ending at some final temperature $T_c \le T < T_i$,
if the rate of change is sufficiently slow the system evolves adiabatically toward its equilibrium state at temperature $T$. (More accurately, we should refer 
to this limit as quasi-static when we are dealing with an open system. We will here use the term adiabatic in the generalized sense.) Small deviations from adiabaticity 
(the quasi-adiabatic regime) can be described by adiabatic perturbation theory (as has been demonstrated explicitly for quenches of quantum systems at
zero temperature,\cite{qaqmc, degrandi_long} and one can anticipate direct analogues for classical quenches). In contrast, if the evolution is fast 
(the quench velocity is high), excitations lead to a large density of defects and the adiabatic description breaks down. The KZ mechanism provides a natural 
way to distinguish these perturbative and non-perturbative regimes. 

According to the arguments of KZ, for the quasi-adiabatic picture to be valid, the 
time $\tau_{\rm q}$ that the system is allowed to take to approach the final temperature $T$ must be at least of the order of the relaxation time $\tau_{\rm rel}$ 
associated with the system's microscopic dynamical properties at that temperature. The relaxation time is simply related to the equilibrium spatial correlation 
length $\xi_T$ according to
\begin{equation}
\label{zdef}
\tau_{\rm rel}  \sim \xi_T^z,
\end{equation}
which defines the dynamic exponent $z$. This exponent depends on the equilibrium universality class of the phase transition, as well as the stochastic dynamics 
imposed on the system (or, alternatively, one can consider Hamiltonian dynamics, e.g., in quantum systems). Thus, for a linear quench with velocity $v$, the 
criterion for staying adiabatic is obtained by requiring for the total quench time $\tau_q$:
\begin{equation}
\label{eq:quench_time}
\tau_{\rm q} \sim |T_i-T|/v \sim \tau_{\rm rel} \sim \xi_T^z \sim |T-T_c|^{-z\nu}, 
\end{equation}
where $\nu$ is the equilibrium correlation-length exponent. 

Another way to interpret the above relationship is to consider the {\it remaining} time $\tau$ of a quench which has reached temperature $T >T_c$ 
after starting out at some $T_i > T$ and which is to continue all the way down to $T_c$. Then, for a given $\tau$, or equivalently, for given velocity $v$,
the relation
\begin{equation}
\label{eq:quench_time_2}
\tau = |T-T_c|/v \sim |T-T_c|^{-z\nu}
\end{equation}
defines the temperature $T$ at which the system falls out of the adiabatic evolution and essentially freezes, 
not being able to evolve significantly for the remainder of the quench process. This should hold independently of the starting temperature $T_i$ if it
is sufficiently above $T$. From this relation we can also extract the velocity (the KZ velocity) 
\begin{equation}
\label{eq:quench_velocity}
v_{\rm KZ}(T) \sim |T-T_c|^{1+z\nu},
\end{equation}
at which the system falls out of adiabaticity at temperature $T$. Thus, it is, in the thermodynamic limit, not possible to stay adiabatic all the
way down to $T_c$. We present an alternative derivation of this result in Appendix \ref{appa}, where we consider the continuous quench as a series
of infinitesimal quenches.

We can also write down the spatial length-scale $\xi_v$ associated with a given velocity, i.e., the correlation length reached at the point
where the infinite system freezes and cannot follow the instantaneous equilibrium state. Since $\xi_v \sim \xi_T$ for the quasi-adiabatic evolution and
$\xi_T \sim |T-T_c|^{-\nu}$  at the point of freezing, Eq.~(\ref{eq:quench_velocity}) gives
\begin{equation}
\label{def:xiv}
\xi_v \sim v^{-1/(z + 1/\nu)}.
\end{equation}
For a finite system the maximum length scale is $L$, i.e., $\xi_v \le L$, and the characteristic velocity separating the adiabatic and non-adiabatic responses 
then has an lower bound, which is simply obtained, according to standard arguments in finite-size scaling theory,\cite{Barber1983} by replacing the largest length-scale 
for the infinite system by $L$. In this case that means $\xi_v \to L$ in (\ref{def:xiv}). Thus, a system of linear size $L$ will remain adiabatic all the way down 
to $T_c$, provided that the quench velocity is  of the order of the size-dependent KZ velocity given by
\begin{equation}
\label{def:vkz}
v_{_{KZ}}(L) \sim L^{-(z+1/\nu)}.
\end{equation}
When the velocity is below this characteristic value, the non-adiabatic response of the system is very small and can be 
treated perturbatively. In contrast, when the velocity exceeds $v_{_{KZ}}(L)$ the quasi-adiabaticity breaks down and the response of the system corresponds 
to non-adiabatic dynamics which is non-perturbative in $v$. 

It should be pointed out that it is in general not possible to assign an exact value to $v_{_{KZ}}(L)$ (and all the other quantities defined above), as 
Eq.~(\ref{def:vkz}) only indicates a proportionality and the change between the quasi-adiabatic and non-perturbative regime normally 
takes place in the form of a smooth cross-over (although we will also demonstrate an interesting exception, where the break-down of the quasi-adiabatic 
regime is sudden). We will here use extensive MC simulations to extract scaling functions of the form $f(v/v_{\rm KZ})$ describing the dynamic approach to 
the critical point for several models and dynamic schemes, from which the cross-over scale can be readily read-off. In addition to the KZ scale, we will also 
investigate and quantify another, higher-velocity (diabatic) cross-over scale $v_a$ related to a size-independent microscopic (lattice) scale $a$.

\subsection{Dynamic exponents}
\label{subsec:z}

As we have seen in the discussion above, the dynamic scaling will naturally involve the dynamic exponent $z$ of a given combination of model and 
imposed MC dynamics (updating scheme for the system configurations). For Metropolis dynamics,\cite{J.Chem.Phys.21.1087} in which $N$ single-spin flip 
attempts define a unit of time in updating the system configurations, many works have been devoted to extracting the value of $z$ (which in the case of
Metropolis dynamics we will often call $z_{\rm M}$) for the 2D Ising model, e.g., 
Refs.~\onlinecite{J.Phys.A.26.6691, PhysRevE.56.2310, Ito1993591, pramana.64.871, PhysRevB.62.1089, jpsj.77.014002, J.Phys.A.23.L453, J.Phys.A.25.2139}. The 
values obtained are typically close to $2.2$, with $z_{\rm M}=2.1667(5)$, obtained in Ref.~\onlinecite{PhysRevB.62.1089}, often quoted as the most reliable 
result. The relatively large dynamic exponent implies that the Metropolis algorithm suffers rather severely from critical slowing-down \cite{jpsj.27.1421} when the 
system is close to its critical point---the collective critical clusters persist for long times when updated only gradually by single-spin flips. 
Despite of the critical slowing-down issue, Metropolis dynamics is still indispensable in its own right due to its close correspondence to relaxation 
processes due to local couplings to the environment in experiment systems.\cite{PhysRevB.71.144406} Moreover, the Metropolis algorithm is very widely
applicable to simulations even of very complex many-body systems. Even though there is no experimental counterpart of cluster updates, efficient cluster 
updates such as the Swendsen-Wang (SW) \cite{PhysRevLett.58.86} and Wolff algorithms \cite{PhysRevLett.62.361} have been very important to reduce or 
eliminate the inefficiency caused by critical slowing down in simulations. However, unlike the Metropolis scheme, the applicability in practice of these 
algorithms is restricted to a smaller number of models.

For SW updates of the 2D Ising model, where the system is subdivided into clusters and each cluster is flipped with probability $1/2$, the nature of the 
dynamic scaling is still somewhat controversial. Values for the dynamic exponent have typically fallen in the range $z_{\rm SW} = 0.2 \sim 0.35$, 
\cite{PhysRevB.43.10617, PhysRevLett.68.962, PhysRevLett.74.3396, PhysRevLett.62.163} but in some works it was instead proposed that the 
characteristic time diverges not as a power $L^z$ but logarithmically, which would imply $z_{\rm SW}=0$.\cite{J.Stat.Mech.2006.P05004} For the 3D
Ising model the exponent is not known very precisely, with results typically falling in the range 
$z_{\rm SW}=0.44 \sim 0.75$.\cite{PhysRevLett.62.163, J.Stat.Phys.58.1083, PhysRevLett.68.962, Ossola.Nucl.Phys.B.691.259}

In the Wolff algorithm, which can be regarded as an improvement over the SW algorithm, clusters are constructed one at a time and always flipped. It is therefore
normally more likely to flip large clusters.\cite{J.Stat.Phys.58.1083} The value of the dynamic exponent was estimated at  $z_{\rm W} \approx 0.3$ for the 2D Ising
model, and in the range  $z_{\rm W} \sim 0.28$ to $0.44$ for the 3D Ising model.\cite{PhysRevLett.68.962, J.Stat.Phys.58.1083} 

\subsection{Aims and outline of the paper}

We here explore dynamic critical scaling in MC simulations of the Ising model, primarily in two dimensions but with some results also 
for higher dimensions. We change the temperature linearly or nonlinearly as a function of MC time and focus on the approach to the critical point.  
When such a quench becomes extremely slow (and perhaps is more properly referred to as annealing), the scheme described above is known 
as \textit{simulated annealing}.\cite{Kirkpatrick13051983} 

While ideas of how to incorporate insights from the KZ mechanism or similar considerations into simulated annealing processes have been discussed previously, 
\cite{J.Stat.Mech.P03032, J.Phys.Conf.Seri.302.012046, Nourani.J.Phys.A.31.8373} the goal of these works has normally been to maximize the efficiency of the process
of finding the global energy minimum of a system (optimizing the annealing schedule), or to reach the finite-temperature equilibrium distribution as 
fast as possible. Also, simulated annealing was studied to analyze the interplay between the KZ mechanism and coarsening dynamics.\cite{PhysRevE.81.050101, J.Stat.Mech.P02014} 
In our work presented here, the objective is instead to study the scaling behavior when the transition point is approached in systems of different size and at different 
velocities. 

The basic idea is to generalize the standard finite-size scaling techniques, where scaling functions depend on the ratio $L/\xi_T$, to finite-velocity 
scaling where $L/\xi_v$ should enter in a similar way. Our main aim is to establish bench-marks for dynamical critical scaling, especially the form of the scaling 
functions describing quenches to $T_c$, for a prototypical model system and the above mentioned most commonly used MC updating schemes. Some aspects of this 
kind of generalized KZ scaling have already been reported, e.g., in quantum systems where similar scaling behavior applies, \cite{EPL.84.67008, PhysRevB.84.224303,  PhysRevLett.109.015701, qaqmc} 
in some classical systems based on effective dynamical Ginzburg-Landau models\cite{PhysRevB.86.064304}, and in the context of soliton formation that can be described by the stochastic Gross-Pitaeveski equation.\cite{PhysRevLett.104.160404}  In Ref.~\onlinecite{Gong10}, Gong \textit{et al.}~studied 
a similar non-equilibrium setup with external-field tuning in infinite-size limit, motivated by general scaling arguments and an RG approach. Some studies
have also been reported of linear temperature quenches similar to those discussed in this paper.\cite{PRB.71.132402,Zhong11} Here we propose different
ways to analyze data and provide a more complete characterization of the scaling behaviors in the entire velocity range. 

We study basic classical Ising models described by the generic Hamiltonian 
\begin{equation}
\label{eq:Hising}
H = -J \sum_{\left\langle i,j \right\rangle} \sigma_i \sigma_j,
\end{equation}
\noindent
where the coupling is ferromagnetic, $J > 0$, and the spins take values $\sigma_i = \pm 1$. The site pairs $\langle i,j \rangle$ normally correspond
to nearest neighbors (and we then impose periodic boundary conditions) but we will also consider the fully-connected model (i.e., all site pairs are 
included in the summation). We discuss the two-dimensional (2D) case in the main results section and discuss the three-dimensional (3D) and fully connected 
cases in Appendix \ref{appb}. For the dynamics, we use single-spin flips accepted according to the Metropolis algorithm \cite{J.Chem.Phys.21.1087} as as well as two 
different cluster algorithms; those of Swendsen-Wang \cite{PhysRevLett.58.86} and Wolff.\cite{PhysRevLett.62.361} 

The rest of the paper is organized as follows: In Sec.~\ref{sec:scaling} we discuss details of the dynamic scaling of the order parameter for linear 
and generalized non-linear power-law protocols through which the system is quenched to the critical point. We also discuss the use of different scaling functions 
applicable in the low-velocity (quasi-adiabatic) and high-velocity (diabatic) regimes, as well as in the regime (a universal scaling regime) connecting these 
behaviors. In Sec.~\ref{sec:results} we demonstrate the application of the dynamic scaling 
ansatz using simulation data obtained with the three different MC updating schemes for the 2D Ising model. In Sec.~\ref{sec:summary} we summarize our main 
conclusions and discuss potential further applications. An alternative derivation of the KZ velocity is provided in Appendix \ref{appa}, where we also
briefly discuss optimized protocols given finite time resources for quenching. In Appendix \ref{appb} we demonstrate dynamic scaling with SW and 
Wolff cluster updates for the 3D and fully-connected Ising models.

\section{Dynamic Finite-Size Scaling}
\label{sec:scaling}

It is well known in equilibrium physics that systems show universal finite-size scaling behavior in the neighborhood of the critical temperature $T_c$. 
Physical quantities can then be described by a non-singular scaling function $g(L/\xi_{_T})$ and a universal power of the system size according to the form 
\begin{equation}
\label{atl}
A(L,T)=L^{\kappa/\nu}g(L/\xi_{_T})=L^{\kappa/\nu}G[(T-T_c) L^{1/\nu}],
\end{equation}
where $\kappa$ is an exponent depending on the universality class of the transition and the quantity $A$.
This general equilibrium form was initially hypothesized based on observations and has now been rigorously demonstrated through the
renormalization group.\cite{RevModPhys.47.773,Barber1983} 

We here discuss how the KZ mechanism introduced in Sec.~\ref{subsec:kz} can be incorporated into finite-size scaling forms for systems undergoing quench dynamics.

\subsection{Generalized KZ finite-size scaling}

In a non-equilibrium setup, which we here first take to be a linear quench toward the critical point, the scaling argument $L/\xi_v$, with $\xi_v$ defined
in Eq.~(\ref{def:xiv}) should enter in addition to the equilibrium argument $L/\xi_T$. Equivalently, as is clear from the definitions in Sec.~\ref{subsec:kz}, we
can also consider the velocity ratio $v/v_{KZ}(L)$. We use it to write down an ansatz in terms of a function depending on the two scaling arguments;
\begin{eqnarray}
A(T,L,v) & = & L^{\kappa/\nu}f(L/\xi_{_T},v/v_{_{KZ}}) \label{atlv} \\
         & = &  L^{\kappa/\nu} F \big[ (T-T_c) L^{1/\nu}, v L^{z+1/\nu} \big]. \nonumber
\end{eqnarray}
This generalized scaling ansatz has been justified in quantum systems in the slow limit using adiabatic perturbation theory,\cite{PhysRevB.72.161201, degrandi_long} 
and it has also been demonstrated in the case of quantum phase transitions in imaginary-time dynamics.\cite{PhysRevB.84.224303, qaqmc} However, except for several
works by Zhong and collaborators (where an $L \to \infty$ formalism was mostly adopted from the outset) \cite{PRB.71.132402,Gong10,Zhong11} and Chandran 
\textit{et al.}~\cite{PhysRevB.86.064304} the classical counterpart has not, to our knowledge, been investigated as extensively as the quantum case. 

\subsection{Linear quench protocol and procedures}

Clearly, Eq.~(\ref{atlv}) reduces to the standard equilibrium finite-size scaling ansatz in the limit $v \rightarrow 0$. When $v \neq 0$ the framework allows us 
to study the response of the system away from the adiabatic limit. For a system with a known value of $T_c$, one can carry out a quench process from a 
high temperature, $T_i>T_c$ to $T_c$, hence eliminating the first argument in the universal function in (\ref{atlv});
\begin{equation}
\label{atclv}
A(T_c,L,v) = L^{\kappa/\nu} F (v L^{z+1/\nu}).
\end{equation}
This scaling form is very similar to the equilibrium form (\ref{atl}) and is easy to study the size and velocity dependence of physical quantities at the transition point,
using data-collapse techniques familiar from conventional finite-size scaling.

The main purpose of the work reported here is to justify 
the generalized scaling ansatz (\ref{atclv}) at $T=T_c$ by testing it in detail for classical phase transitions and investigating its range of applicability. We present 
several bench-mark cases showing that the ansatz works extremely well. Below we will also extend Eq.~(\ref{atclv}) by introducing yet another scaling argument 
$v/v_a$, where $v_a$ is related to a microscopic scale. One can then observe scaling over the entire velocity range $v \in [0,\infty]$.

Theoretically, any temperature higher than $T_c$ can be used as the initial temperature (or one can start below $T_c$ from an ordered state, but
here we will only consider $T_i > T_c$) but in practice a higher temperature implies that it is easier to generate an equilibrated configuration before the quench 
process begins (which would be particularly important when studying spin glasses or related systems with very slow equilibration close to $T_c$). The details 
of the diabatic dynamics will also of course depend on $T_i$, but for slower velocities the results should become independent of the initial condition.

Knowing the exact value of $T_c$ prior to the simulation is not a necessary condition for this approach to work, since one can also track, e.g., the order parameter 
or the Binder cumulant \cite{PhysRevLett.47.693} in non-equilibrium simulations and locate $T_c$ by various scaling techniques similar to equilibrium finite-size scaling. 
We demonstrated this recently for a quantum model.\cite{qaqmc} However, for purpose of demonstrating  dynamic scaling at classical transitions under different dynamic schemes, we will here use the known values of $T_c$ for the systems of interest. 

\begin{figure}
  \includegraphics[width=7.5cm, clip=true]{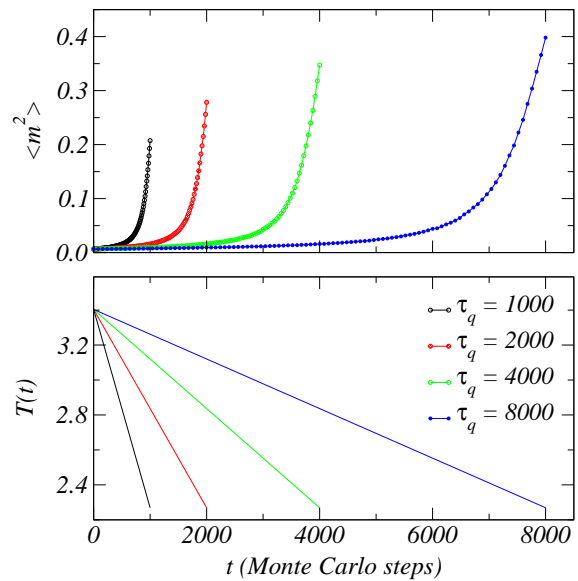}
  \caption{ (Color online) Illustration of linear quenches of the 2D Ising model. A system of size $N=32 \times 32$ was equilibrated at the initial temperature 
$T_i = 1.5 T_c$ and was thereafter linearly quenched to $T_c$. The quench velocity was $v=0.5 T_c/\tau_q$, where $\tau_q$ is the total quench time. Here one unit of time 
is defined as one MC step consisting $N$ attempts to flip randomly selected spins using the standard Metropolis probability. Shown are the temperature (bottom panel) 
and the magnetization squared (top panel) versus time for different total quench times. We will focus our studies here on the scaling of $\langle m^2\rangle$
at the final point.}
    \label{fig01}
\end{figure}
 
For obtaining the results presented in this paper, we typically started with an equilibrated configuration at an initial temperature $T_i = 1.5 T_c$ and performed an MC 
quench process carrying the system to its critical point. The quench velocity in the linear case can therefore be written as $v = 0.5T_c/\tau_q$, where $\tau_q$ 
is the total quench time in units of MC steps. We note that one unit of time in MC simulations normally corresponds to an extensive number of spin flips 
(but we will also consider a case, with Wolff dynamics, where this is not true). The precise definition of the time unit depends on the dynamics used. 

Typical examples of linear quench processes are illustrated in Fig.~\ref{fig01} with results for the magnetization squared (which will be the only physical
quantity studied in this paper);
\begin{equation}
\label{eq:m2def}
m^2 = \left ( \frac{1}{N} \sum_{i=1}^N \sigma_i \right )^2.
\end{equation} 
In this case the exponent $\kappa=-2\beta$ in Eq.~(\ref{atclv}) and we expect scaling at $T_c$ according to
\begin{equation}
\label{eq:m2scaling1}
\langle m^2 \rangle = L^{-2 \beta/\nu} F \big( v L^{z +1/\nu} \big),
\end{equation} 
provided $v$ is sufficiently small (and we will discuss how small that is below). Note that the process stops at $T_c$ and there is no waiting
time after that to relax the system further (which would introduce yet another time scale, which one can certainly consider but we do not include it here).
Only a single measurement of $m^2$ is carried out after the system has reached $T_c$ and the brackets $\langle \dots \rangle$ in Eq.~(\ref{eq:m2scaling1})
represent the ensemble average over different quenches with different equilibrated starting configurations.
Typically, we calculated averages on the basis of thousands of such independent MC runs. 

The initial configuration at $T=T_i$ was equilibrated and sampled before the start of each run using cluster updates (to be discussed further below)  to ensure 
statistically independent starting configurations for each quench process. For studying slow dynamics it is strictly not necessary to equilibrate the initial 
configuration, since one can expect the system to become memoryless for slow enough quenches when approaching $T_c$. However, we here also study 
the fast limit and want the system to reduce to the equilibrium at $T_i$ when $v \to \infty$. We therefore always equilibrate.

\subsection{Nonlinear quench protocols}

The simple scaling hypothesis discussed above has also been generalized to non-linear protocols, where the critical point is approached according to an arbitrary 
power-law of the time $t$ measured with respect to the final time $\tau_q$,\cite{PhysRevB.81.012303, v_comment}
\begin{equation}
\label{rquench}
T-T_c = v(\tau_q-t)^r,
\end{equation}
where $v$ is the velocity as above for a linear quench ($r=1$), the acceleration (up to a factor $2$)
for a quadratic quench ($r=2$), etc.~(and for simplicity we will refer to $v$ as the velocity, regardless of the power $r$). For a sudden quench ($r=0$) $v$ should be regarded as the amplitude of the change in $T$ (and with this definition note that there is no waiting time after the quench, which is another time-scale that could be added but we do not consider here). As in the linear case, for all $r$ we use $T_i = 1.5 T_c$ and express $v$ in units of $T_c$ as $v = 0.5T_c/\tau_q^r$, where $\tau_q$ is the total quench time.

The generalized critical ``velocity'' for arbitrary $r$ (including non-integer) can be easily found by following the same arguments as in Sec.~\ref{subsec:kz};
\begin{equation}
\label{eq:vkz}
v_{_{KZ}}(L) \sim L^{-(z r + 1/\nu)}.
\end{equation}
In Appendix \ref{appa} we provide an alternative derivation of this result based on a time-discretized quench, which also gives some information on how the
unknown prefactor above depends on the exponents involved. 

The magnetization scaling form (\ref{eq:m2scaling1}) with (\ref{eq:vkz}) becomes
\begin{equation}
\label{eq:m2scaling}
\langle m^2 \rangle = L^{-2 \beta/\nu} F \big( v L^{zr +1/\nu} \big).
\end{equation} 
We will study mainly $r=1$ quenches but also discuss some results for $r=1/2$ and $r=2$. Protocols for approaching the critical point very slowly, in 
particular with negative $r$, have also been investigated recently.\cite{PhysRevB.86.064304}

\subsection{Complete scaling form for the order parameter}

For a given system size we can access a wide range of velocities (the highest $v <\infty$ corresponding to carrying out a single MC step) and we can therefore examine
very different response regimes of the system. When the quench velocity becomes very high, our procedures ensure that the magnetization squared after the quench 
to $T_c$ remains close to its value at the initial temperature $T_i$. Since the correlation length has a finite value there, one expects, for sufficiently large $L$,
\begin{equation}
    \label{eq:m2_at_high_v}
    \langle m^2 \rangle = \frac{1}{N^2} \sum_{\langle i,j \rangle} \langle \sigma_i \sigma_j \rangle 
  = \frac{1}{N} \sum_{j} \left\langle \sigma_0 \sigma_j \right\rangle  \sim L^{-d},
\end{equation}
where $d$ is the number of dimensions; here $d=2$ except in Appendix \ref{appb}, where we also consider $d=3$ and infinite dimensionality (in which
case $L$ is defined by $L=N^{1/d}$ with $d$ the upper critical dimension).
Thus, in the high-velocity limit, $\langle m^2 \rangle$ should depend on the initial temperature $T_i$ and scale as $L^{-d}$. 

When the velocity decreases one can expect the order of the system to develop gradually, and as long as the KZ correlation length $\xi_v$ is much smaller than the system size $L$ the magnetization squared should still depend on $L$ with the trivial power above. With the scaling form (\ref{eq:m2scaling}), this behavior necessarily implies that the function $F$ in this regime must reduce to a power law of the argument $v L^{z+1/\nu}$;
\begin{equation}
    \label{eq:high_v_power_law}
    \langle m^2 \rangle \sim  L^{-2 \beta/\nu } \big( v L^{zr+1/\nu} \big)^{-x},
\end{equation}
and this exponent can be obtained by demanding this to be proportional to $L^{-d}$, i.e.,
\begin{equation}
   \label{eq:power}
   x = \dfrac{ d - 2 \beta/\nu  }{ zr + 1/\nu }.
\end{equation}
Thus, there is an intermediate universal scaling regime where
\begin{equation}
   \label{eq:m2vx}
    \langle m^2 \rangle \sim  L^{-d}v^{-x}.
\end{equation}
Note that this is not consistent with the high-velocity limit for fixed $L$, where, as discussed above, $\langle m^2 \rangle$ must converge to a constant times
$L^{-d}$ (without any remaining $v$ dependence). The power law written as Eq.~(\ref{eq:high_v_power_law}) should instead hold for arbitrary large values of 
$v L^{z+1/\nu}$, as long as $L$ is sufficiently large. Below we will discuss in detail the cross-overs between the power law and the ultimate high-$v$ 
limit for any $L$.

Finally, when the velocity is decreased further and approaches $v_{_{KZ}}(L)$, the assumption $\xi_v \ll L$ no longer holds. One would then expect deviations 
from the power-law form and a cross-over to a regime where Eq.~(\ref{eq:m2scaling}) tends toward the corresponding $L$-dependent equilibrium value of
at $T_c$, i.e., the standard finite-size behavior scaling, 
\begin{equation}
   \label{eq:fsscale}
\langle m^2\rangle \sim L^{-2\beta/\nu},
\end{equation}
sets in. This cross-over from the $v$-dependent power-law to this equilibrium form is smooth and 
contained in the function $F$ in Eq.~(\ref{eq:m2scaling}).

To incorporate all these different asymptotics in different velocity regimes, it is useful to introduce a short-range length scale $a$, which is of the order of 
one lattice spacing, and, therefore, can be set to 1 for any practical purpose. This non-trivial factor $a$ is essential for defining the 
\textit{engineering dimension},\cite{qpt_sachdev} $a^{-d+2\beta/\nu}$, which compensates for the discrepancy between the scaling dimension $L^{-2\beta/\nu}$ of 
$\langle m^2 \rangle$ and its canonical dimension $L^{-d}$. The short-range length scale sets the size-independent upper limit $v \sim v_a$ beyond 
which the power-law behavior (\ref{eq:m2vx}) should break down;
\begin{equation}
 \label{eq:vkz_a}
 v_a \sim a^{-(zr+1/\nu)}.
\end{equation}
More explicitly, based on the above discussion one cannot expect Eq.~(\ref{eq:m2scaling}) to be able to describe all situations with a single scaling
function $F$, and this function should actually be replaced by two different scaling functions in different regimes of $(v,L)$, namely,
\begin{equation}
 \label{eq:2scaling_functions}
  \langle m^2 \rangle = \left \{
\begin{array}{l c}
L^{-2\beta/\nu} a^{-d+2\beta/\nu} \hspace{1pt} f_1 ( v L^{zr+1/\nu} ),  &  v < v_a \\
& \\
L^{-d} \hspace{1pt} f_2 (a^{-(zr+1/\nu)}v^{-1} ),  & v > v_{{KZ}}(L),
\end{array}
\right.
\end{equation}

\noindent
where $f_1$ and $f_2$ are different scaling functions, valid in their own associated velocity regions. More generally, the above two scaling functions can be described 
by a single common universal form with two arguments, i.e.,
\begin{equation}
    \label{eq:m2gscaling}
    \langle m^2 \rangle \sim  L^{-2 \beta/\nu }a^{-d+2\beta/\nu}G (v L^{zr+1/\nu}, a^{-(zr+1/\nu)} /v).
\end{equation}
However, it is in practice easier to analyze its two limiting forms (\ref{eq:2scaling_functions}) with single scaling arguments.

In the velocity regime $v \ll v_{_{KZ}}(L)$, the system should be perturbative 
in $v$, while in the opposite limit when $v \gg v_a$ the system can be described by perturbation in $1/v$. As we will demonstrate below with numerical data, there 
is a wide region, $ v_{{KZ}}(L) < v < v_a$, over which $f_1$ and $f_2$ are both applicable. This corresponds to the regime where both perturbative descriptions 
(in $v$ and $1/v$) have broken down and have been replaced by a universal power-law behavior, expressed as Eq.~(\ref{eq:high_v_power_law}) and (\ref{eq:m2vx}) 
for $f_1$ and $f_2$, respectively.

The basic idea that we are pursuing throughout this paper is that by quenching the system with different velocities (or generalized velocity for $r \neq 1$), one 
can generally observe cross-over behaviors at $v \sim v_{{KZ}}(L)$ as well as at $v \sim v_a$ between perturbative and non-perturbative regimes. The velocities 
$v_{{KZ}}(L)$ and $v_a$ separate different forms of the size dependencies of the magnetization squared (which is the quantity we focus on here, but one of
course expects analogous behaviors in other quantities). The characteristic velocity $v_a$ separates the velocity independence, $\langle m^2 \rangle \sim  L^{-d}$, 
from the power-law form $\langle m^2 \rangle \sim  L^{-d}v^{-x}$ for $v_{{KZ}}(L) < v < v_a$, and then another characteristic velocity $v_{{KZ}}(L)$ separates 
this behavior from the critical equilibrium scaling form $\langle m^2 \rangle \sim L^{-2 \beta/\nu}$ for $v<v_{{KZ}}(L)$.

The above forms Eq.~(\ref{eq:2scaling_functions}) can be used to analyze numerical data by dividing $\langle m^2 \rangle$ by the appropriate power of $L$ appearing
on the right-hand side and graphing the result versus the argument of the scaling function. The data should then collapse onto the scaling function in the
region of $(v,L)$ where it holds; hence the scaling function is obtained. The first scaling form $f_1$, which requires the knowledge of critical exponents, is analogous to the equilibrium scaling at $T_c$. The second scaling $f_2$, requires no knowledge of the critical exponents. 

Although it is not necessary, we can also assume that the function $f_1$ 
in Eq.~(\ref{eq:2scaling_functions}) can be written as a series expansion of $vL^{z+1/\nu}$ in its perturbative regime, and, as was pointed out above, $f_2$ 
should depend on $T_i$ and can be written as a series expansion in $1/v$ in its perturbative regime. In their non-perturbative regimes both functions 
reduce to the same power law form (just expressed in two different ways). We therefore expect the following forms to hold
in the three different scaling regimes:
\begin{equation}
 \label{eq:velocity_regimes}
  \langle m^2 \rangle = \left \{
\begin{array}{l r}
L^{-2\beta/\nu} \sum \limits_{n} c_n (v L^{zr+1/\nu})^n,& v \lesssim v_{_{KZ}}(L) \\
\\
L^{-d} \left (\frac{1}{v}\right)^x,  & v_{{KZ}}(L)\ll v \ll 1 \\ 
\\
L^{-d} \sum \limits_{n} c_n (1/v)^n,   &  v \gtrsim 1 
\end{array}
\right.
\end{equation}
\noindent
where we have explicitly set $a=1$ and, therefore, $v_a = 1$. In the following, we will refer to the velocity regime $v \lesssim v_{{KZ}}(L)$ as the 
quasi-adiabatic regime, $v_{{KZ}}(L)\ll v \ll 1$ as the universal scaling regime, and $v \gtrsim 1$ as the diabatic regime. The asymptotic form in the universal 
scaling regime $v_{{KZ}}(L)\ll v \ll 1$ corresponds to the power-law behavior, Eq.~(\ref{eq:m2vx}), that both scaling functions $f_1$ and $f_2$ converge to. 
Note again that, in practice, the highest velocity in our simulations corresponds to one MC step, i.e., $v=(T_i-T_c)/\tau_q^r$ with $\tau_q=1$, which is of the order 
$1$ with our chosen initial temperature.

Normally the cross-overs between the different regimes in Eq.~(\ref{eq:velocity_regimes}) are completely smooth, which we will demonstrate in the next section 
for Metropolis and SW dynamics in the 2D Ising model. Remarkably, however, in the case of the Wolff cluster algorithm we will show that the power-law regime is
absent and the cross-over between the two perturbative regimes is not smooth. Instead, in the thermodynamic limit both the quasi-adiabatic and diabatic scaling 
behaviors break down discontinuously at specific values of the scaled velocity. In this sense the Ising model with Wolff dynamics undergoes a dynamic phase 
transition.

\section{Simulation Results}
\label{sec:results}

In this section we demonstrate the application of dynamic finite-size scaling using the standard 2D Ising model on the square lattice. 
We discuss results using Metropolis dynamics in Sec.~\ref{subsec:metropolis}, SW dynamics in Sec.~\ref{subsec:sw}, and Wolff dynamics 
in Sec.~\ref{subsec:wolff}. The exact value of $T_c$ and the critical exponents are known exactly from the Onsager solution:\cite{PhysRev.65.117} 
$T_c/J = 2/\ln(1+\sqrt{2})$, $\nu=1$ and $\beta=1/8$. This system therefore provides a good testing ground for our techniques.  
For all the quench processes we consider in the following we start with an initial temperature $T_i = 1.5 T_c$, using the value of $T_c$ 
quoted above, and then quench the system exactly to the critical point, at which observables are computed (and note again that there is no further waiting
at $T_c$; a single measurement of $\langle m^2\rangle$ is obtained after each quench). The quench process for given parameters is repeated thousands of 
times with different equilibrated starting configurations in order to obtain statistically precise averages. 

The focus here will be how the system responds to the dynamics when crossing the two characteristic velocities defined in the previous section, $v_a$ and $v_{_{KZ}}(L)$ , 
and how the cross-over behaviors emerge in the dynamic scaling.  As shown explicitly in the scaling forms discussed above, the scaling naturally involves the dynamic 
exponent $z$. Since the 2D Ising equilibrium critical exponents are all known exactly, the dynamic scaling allows one to extract $z$ independently (and note
that this exponent depends on the dynamics imposed and is not known exactly for any of the schemes we use). In practice we here do this by optimizing a data collapse 
(onto one of the unknown scaling functions, which the process yields) with $z$ as the only adjustable parameter.

\subsection{Metropolis dynamics}
\label{subsec:metropolis}

Typical linear quench processes using Metropolis dynamics have been shown in Fig.~\ref{fig01}. We here follow the convention that 
one unit of time is defined as $N=L^2$ attempts of flipping a randomly selected spin with the acceptance probability $p={\rm min}[1,{\rm e}^{-\Delta E/T}]$,
where $\Delta E$ is the change in energy after flipping the spin. For convenience we will give velocities in units of $T_c$, i.e., with the above $T_i$ we 
define $v=0.5/\tau_q^r$ for total quench time $\tau_q$ in units of MC steps. To demonstrate the insensitivity of the scaling to $T_i$, we will also
present a test of this assumption. We first discuss the linear quenches and then present some results also for $r=1/2$ and $r=2$ non-linear protocols 
according to Eq.~(\ref{rquench}). 

\subsubsection{Linear quench}

\begin{figure}
   \includegraphics[width=8.4cm,clip=true]{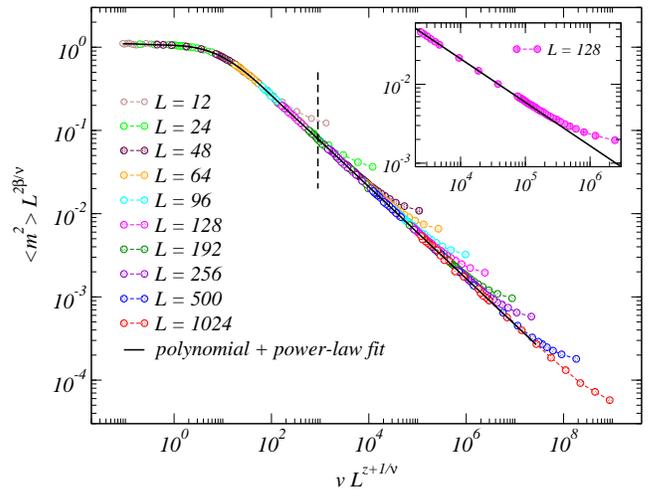}
   \caption{(Color online) The squared magnetization scaled by $L^{2\beta/\nu}$ after linear quenches to $T_c$, using Metropolis dynamics for 2D 
             Ising models of different sizes. The collapsed data correspond to the first scaling form, $f_1$, in Eq.~(\ref{eq:2scaling_functions}). The expected 
             three different regimes corresponding to different asymptotics can be clearly distinguished; (left) approach to the equilibrium critical scaling in the 
             quasi-adiabatic regime $v \lesssim v_{{KZ}}(L)$, (center) power-law scaling in the universal regime $v_{_{KZ}}(L)\ll v \ll 1$, and (right-most points for 
             each $L$) deviations from the scaling function in the diabatic regime $v\gtrsim 1$. The vertical dashed line shows the point separating the two fitting 
             windows used in the optimization of the data collapse (varying $z$); to the left the fitted function approximating $f_1$ in the quasi-adiabatic 
             regime is a high-order polynomial, and to the right a pure power 
             law (straight line) given by Eq.~(\ref{eq:power}) is used to account for the universal scaling behavior. The diabatic tails for each $L$ were not 
             included in the fits. The dynamic exponent used in scaling the $x$-axis was adjusted to obtain the overall best simultaneous fits of the data 
             in the quasi-adiabatic and scaling regimes, which resulted in $z_{M}=2.172(3)$ with the goodness of the fit, $\chi^2/{\rm dof} \approx 1.0$. 
             Error bars for the data points are all smaller than the symbol sizes. The inset shows details of the $L=128$ data in the region where the behavior
             crosses over from universal scaling to diabatic.} 
   \label{fig02}
\end{figure}

\begin{figure}
   \includegraphics[width=8.4cm,clip=true]{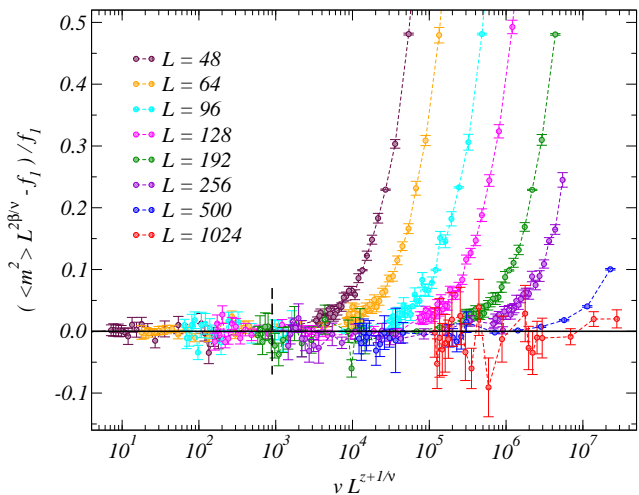}
   \caption{(Color online) The difference between the fitted function $f_1 (vL^{z+1/\nu})$ and the scaled magnetization squared, 
            $\langle m^2 \rangle L^{2\beta/\nu}$ (same data as in Fig.~\ref{fig01}). The vertical line shows the point separating the two fitting windows. 
            The points deviating significantly from the horizontal line correspond to diabatic behavior and those points were systematically excluded 
            in the fitting procedure.}
   \label{fig03}
\end{figure}

\begin{figure}
   \includegraphics[width=8.4cm,clip=true]{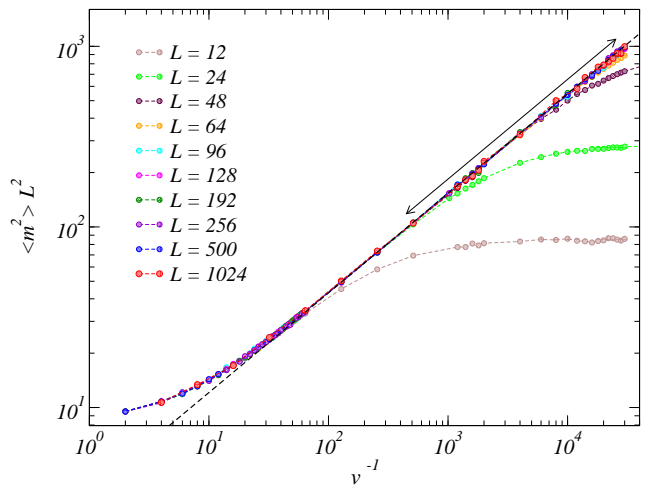}
   \caption{(Color online) Data collapse producing the second scaling function, $f_2$, in Eq.~(\ref{eq:2scaling_functions}) for linear Metropolis-quenches with different system sizes. Here the left region corresponds to the diabatic regime, while straight-line form corresponds to the universal power-law scaling regime. The points deviate from the common function $f_2$ in the $L$-dependent quasi-adiabatic regime. The dashed line shows the slope expected with the dynamic exponent extracted in Fig.~\ref{fig02}. As will be discussed in Sec.~\ref{sec:combination}, the line with arrow indicates the region selected for the linear fit after taking log-log, we consider sizes $L \ge 192$. The linear fit gives the slope $x_{r_1} = 0.550(3)$, with $\chi^2/\text{dof} \approx 1.0$, which implies $z_M=2.17(1)$, consistent with the result obtained in Fig.~\ref{fig02} of $f_1$ scaling.}     
   \label{fig04}
\end{figure}

\begin{figure}
   \includegraphics[width=8.4cm,clip=true]{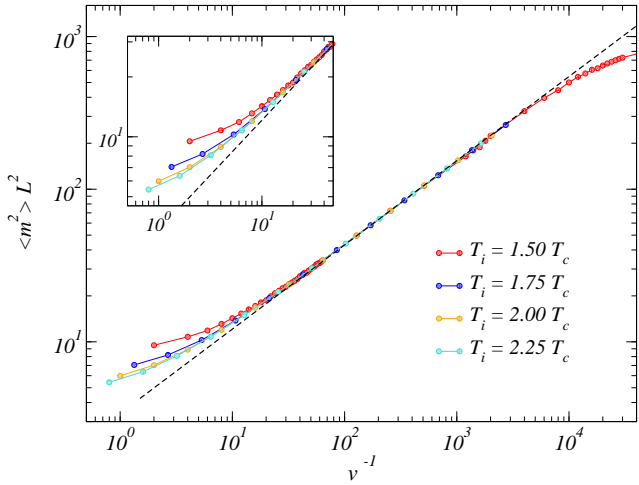}
   \caption{(Color online) Data graphed according to the second scaling function, $f_2$, in Eq.~(\ref{eq:2scaling_functions}) for linear Metropolis quenches on $L=48$ system with different initial temperatures. As expected, $f_2$ nominally depends on the initial temperature only in the diabatic regime. The inset shows more details of the data in the region where the data converge to  common curve.}             
   \label{fig05}
\end{figure}

Data sets for different system sizes in linear-quench simulations at different velocities are analyzed collectively in Fig.~\ref{fig02}, using the scaling 
procedure appropriate when the first scaling form in Eq.~(\ref{eq:2scaling_functions}) applies. Scaling collapse giving the function $f_1$ is observed all the way from the 
adiabatic regime, crossing over into universal power-law scaling, which persists up to arbitrarily large values of the KZ scaled velocity $vL^{z+1/\nu}$ when increasingly
large $L$ is used (pushing the diabatic cross-over further to the right). As we discussed in Sec.~\ref{sec:scaling}, the scaling behavior allows one to 
determine the dynamic exponent by carrying out a fitting procedure in which the value of $z$ is adjusted to give the optimal fit to all the data included, 
which we quantify using the standard $\chi^2$ per degree of freedom (dof). We here use two different functional forms to describe the function $f_1$ in the
fitting procedure, as in the first two lines of Eq.~(\ref {eq:velocity_regimes}), in two non-overlapping windows of the scaling argument $vL^{z+1/\nu}$.
The ``plateau and shoulder'' in Fig.~\ref{fig02} correspond to the quasi-adiabatic regime $v \lesssim v_{KZ}(L)$ and we use a high-order polynomial 
fit in this window. For practical purposes, to minimize the order of the polynomial required, we fit to the log-log data instead of the original data.
The second window corresponds to the universal scaling regime characterized by $v_{{KZ}}(L)\ll v \ll 1$, where we use a pure power-law fit (a straight-line 
fit on the log-log scale).

The point separating the two fitting windows is chosen such that $\chi^2$ computed individually in each window is statistically good. High-$v$ data 
are excluded for each $L$ when they deviate from the common curve (in which case they will also ruin the goodness of the fit, thus allowing for systematic 
exclusion of diabatic data). There are of course scaling corrections expected, as in standard finite-size scaling of equilibrium data, but for the Ising model 
these are relatively small.\cite{sandvik10} We obtain good fits by considering system sizes $L \ge 12$. The statistical error of $z$ is computed by repeating
the data-collapse procedure many times with Gaussian noise added to the MC data (with standard deviation given by the corresponding error bars of the data).
This procedure gives $z_{M}=2.172(3)$, which is in good agreement with values previously obtained, e.g., 
in Refs.~\onlinecite{Ito1993591, PhysRevE.56.2310, pramana.64.871, PhysRevB.62.1089}.

Fig.~\ref{fig03} further illustrates the optimization procedure in the form of the 
deviation of the two-piece fitting function $f_1(vL^{z+1/\nu})$ from the MC data for $\langle m^2 \rangle L^{2\beta/\nu}$. Statistically, the data points fully 
obey the scaling collapse except for those corresponding to the diabatic $v\gtrsim 1$ regime (which are excluded from the fits). 

Note that the purpose of parametrizing the scaling function and carrying out fits is only to provide a convenient way to define the goodness of the 
data collapse. As long as the imposed functional form is capable of reproducing the scaling function to within the precision set by the error bars of
the data (which is self-consistently tested by the statistical soundness of the fit quantified by $\chi^2$), this procedure in no way distorts or biases 
the data collapse.

We discuss the diabatic regime next. Data collapse according to the second of Eqs.~(\ref{eq:2scaling_functions}) is shown in Fig.~\ref{fig04}. The dashed line in Fig~\ref{fig04} is drawn according the the value of $x$ given by the result of $z$ from  Fig~\ref{fig02}. However, independently, the power-law behavior corresponding to the straight line allows one to estimate the dynamic exponent in a straight-forward manner through this kind of analysis, given the relation between the power (slope) $x$ and $z$ in Eq.~(\ref{eq:power}). The advantage of this procedure is that the rescaling of the data does not involve any critical exponents at all. As we will discuss in more detail in Sec.~\ref{sec:combination}, using linear fit after taking log-log, we obtain $x_{r_1}=0.550(3)$, which implies $z_M = 2.17(1)$, this is consistent with the result obtained by $f_1$ scaling.

Note again that the linear regions in Figs.~\ref{fig02} and \ref{fig04} correspond to the same 
data points, falling within the universal scaling regime, with just two different ways of expressing the middle line of Eq.~(\ref{eq:velocity_regimes}),
as stated according to Eq.~(\ref{eq:m2vx}) or as in Eq.~(\ref{eq:high_v_power_law}) by moving the appropriate power of $L$ to the left.

As discussed in Sec.~\ref{sec:scaling}, the initial temperature $T_i$ at which the system is equilibrated before the quench process begins should only have 
a nominal effect on the scaling. We normally use $T_i=1.5T_c$, but to demonstrate the insensitivity of the scaling to the initial temperature we show
in Fig.~\ref{fig05} results for a fixed system size and several values of $T_i$. As expected, there are differences in the diabatic regime, where in the
$v \to \infty$ limit the results converge to the equilibrium at $T_i$. Beyond this regime, at lower velocities the data quickly converge to a common curve in 
the universal scaling regime. The convergence to a pure power law is somewhat faster for higher $T_i$, but it should be noted that the simulation time increases
with $T_i$, which implies that, for purposes of extracting the dynamic exponent by fitting a straight line, there is some trade-off between the faster
convergence to the power law and the longer simulation time. In cases where the initial equilibration may be challenging close to $T_c$, e.g., in
frustrated systems (especially glasses) where cluster algorithms cannot be used, one may also want to start at a high $T_i$ in order to ensure good
initial equilibration.

\subsubsection{Non-linear quenches}

\begin{figure}
   \includegraphics[width=7.4cm,clip=true]{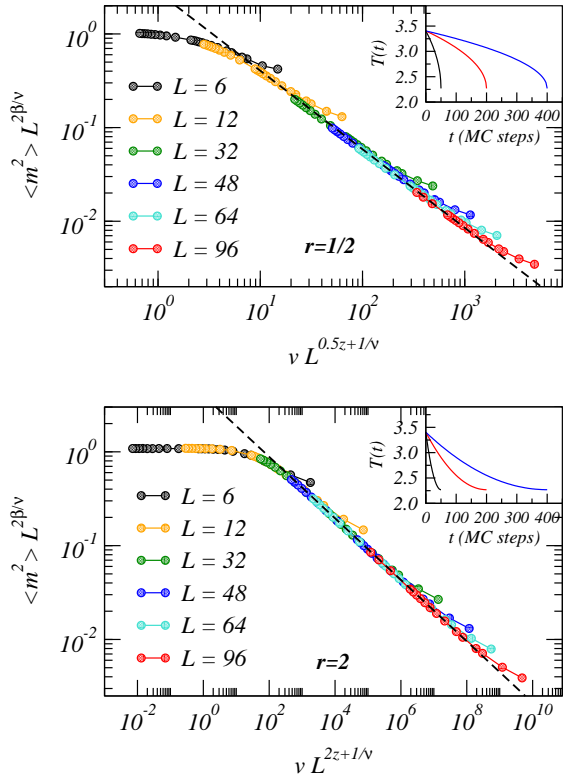}
   \caption{ (Color online) Data collapse in non-linear quenches to $T_c$ with Metropolis dynamics of the 2D Ising model. The top and bottom panels are for $r=1/2$ and $r=2$, respectively, and $v$ is expressed in units of $T_c$ according to the definition (\ref{rquench}) of the protocols with $T_i=1.5T_c$. The data are analyzed and graphed in the same way as the linear quenches in Fig.~\ref{fig02}. The dashed lines show the slopes expected according to Eq.~(\ref{eq:power}) with the dynamic exponent extracted in Fig.~\ref{fig02}. The insets shows examples of the protocols used.}
   \label{fig06}
\end{figure}

\begin{figure}
   \includegraphics[width=7.4cm,clip=true]{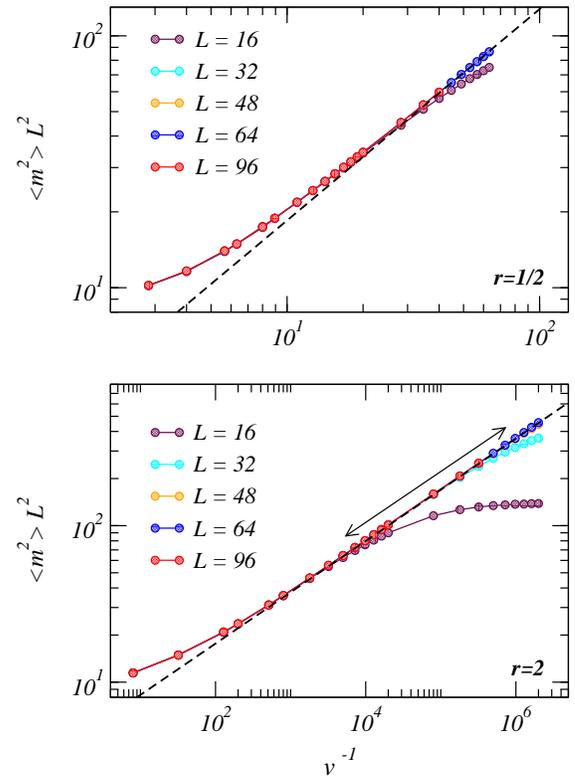}
   \caption{ (Color online) Data collapse in the diabatic and universal scaling regimes for non-linear Metropolis quenches, with the top and bottom panels for protocols
             with $r=1/2$ and $r=2$, respectively. The dashed lines show the slopes expected according to Eq.~(\ref{eq:power}) with the dynamic 
             exponent extracted in Fig.~\ref{fig02}. As we will discuss in Sec.~\ref{sec:combination}, the line in $r=2$ panel indicates the region selected for linear fit after taking log-log, we consider sizes $L \ge 48$. The linear fit yields the slope $x_{r_2} = 0.32689(7)$, with $\chi^2/\text{dof} \approx 0.9$, which implies $z_M=2.1767(5)$, consistent with the result obtained in Fig.~\ref{fig02}.} 
   \label{fig07}
\end{figure}

As we pointed out in Sec.~\ref{sec:scaling}, the KZ scaling scheme is not restricted to linear protocols. Eq.~(\ref{eq:2scaling_functions}) incorporate nonlinear quench scenarios through the exponent $r$ in the definition (\ref{rquench}) of the more general protocol. These expressions provide a simple way to separate the quench process from the underlying stochastic dynamics (updating scheme); the former is characterized by the parameter $r$ and the latter by the dynamic exponent $z$. Since in the non-equilibrium scaling relations only the combination $zr$ enters, changing the exponent $r$ has an effect similar to manipulating  the dynamical exponent, which potentially can be useful for optimization and other purposes. Here we just focus on testing the applicability of scaling with $r\not=1$.

In Fig.~\ref{fig06}, bottom panel, we show results of a ``constant acceleration'' quadratic quench with $r=2$, in which case the the characteristic quantity (\ref{eq:vkz}) 
stands for a critical acceleration  separating a perturbative and non-perturbative regimes, in analogy with the earlier discussion of the linear quench. The scaling collapse
works very well, apart from the expected ultra-high acceleration limit where a break-down again is expected (and the deviations can be analyzed in terms
of a different scaling function, as we will do below). 
The dynamic scaling also holds when the parameter $r$ is a non-integer number corresponding to a non-analytic protocol, as it should, based on the derivations of the
KZ scaling in Appendix \ref{appa} (while the applicability for non-integer $r$ is less clear from other generalizations considered for 
$r \not=1$ \cite{PhysRevB.81.012303, v_comment}). We demonstrate this with results of a square-root quench, $r=1/2$, in the top panel of Fig.~\ref{fig06}.

As shown in Fig.~\ref{fig07}, scaling collapse also works very well in the diabatic limit of both these nonlinear quench protocols. A cross-over of the function $f_2$ to the universal scaling regime is observed as in the $r=1$ case. Most importantly, the power-law behavior is clearly observed. One can again use the power-law regime in the $f_2$ scaling to estimate the power $x$, which can then be translated to $z$ according to Eq.~(\ref{eq:power}). Using linear fit after taking log-log in the $f_2$ scaling of $r=2$ quench, we obtain the slope 
$x_{r_2} = 0.32689(7)$, which implies $z_M = 2.1767(5)$. Remarkably, the statistical precision of this result is higher than the best results based on the $r=1$ quenches discussed 
above, although the system sizes there were considerably larger. It is then interesting to ask what the optimal $r$ is for extracting $z_M$, but we have not investigated 
this systematically. The applicability of the dual scaling for arbitrary $r$ also opens an interesting opportunity to independently extract all of the exponents 
$\beta$, $\nu$, and $z$, as we will discuss in Sec.~\ref{sec:combination}.

\subsubsection{Combining results from different quenches}
\label{sec:combination}

As shown in the previous section, the $f_1$ and $f_2$ dual scaling behaviors are clearly observed in both linear and non-linear quenches. The $f_2$ scaling scheme is particularly interesting in practice. As we mentioned in Sec.~\ref{sec:scaling}, $f_2$ scaling does not involve the prior knowledge or optimization of the critical exponents, 
while the power in the universal scaling regime still carries the information of the critical exponents through Eq.~(\ref{eq:power}). The power $x$ can be measured easily by linear log-log fit of the data. This property of dual scaling and the convenient way of extracting the power $x$ from $f_2$ scaling for any arbitrary $r$ open an interesting opportunity to extract the exponents $z$, $\nu$, and $\beta$ in a completely independent way.

We first point out some important aspects of the applications of $f_2$ scaling. One important aspect of $f_2$ scaling is that it corresponds to the regime in which the correlation length $\xi_v$ Eq.~(\ref{def:xiv}) is growing as $v$ decreases, while $\xi_v$ is still much smaller than the system size, i.e., $\xi_v \ll L$. Effectively, in this regime the rescaled quantity $\langle m^2 \rangle L^2$ is size-independent. This property provides a simple way to do the linear fit in practice: one can simply follow the largest available sizes, when the data points from these sizes become indistinguishable in the $f_2$ plot and the system sizes, thus, are large enough to be effectively free of finite-size effects. This $L$ convergence aspect is seen in Figs.~\ref{fig04} and \ref{fig07}. Quantitatively, one can again use $\chi^2/\text{dof}$ to quantify the result. If small sizes that potentially carry finite-size effect are included in the linear fit, they will certainly ruin the goodness of the statistics. The same principle can also be used for selecting the region for linear fit. The ideal region for the linear fit should be in the power-law regime. If the data points from either the quasi-adiabatic or diabatic regime are included, they will also ruin the goodness of the fit quantified by $\chi^2/\text{dof}$. 

In the following we use $r=1$ and $r=2$ quenches to demonstrate the idea outlined above. For $r=1$ quench as shown in Fig.~\ref{fig04}, we consider sizes $L \ge 192$ since the data points from these sizes already show indistinguishable behavior in the power-law regime. The region in which linear fit is performed is indicated by the line with arrows. The selection of the region is determined by the minimization of $\chi^2/\text{dof}$. We obtain $x_{r_1} = 0.550(3)$ with $\chi^2/\text{dof} \approx 1.0$. For $r=2$ quench shown in the bottom panel of Fig.~\ref{fig07}, using the same principle for selecting the system sizes (with $L \ge 48$) and the region for linear fit (indicated by the line with arrows), we obtain $x_{r_2} = 0.32689(7)$ with $\chi^2/\text{dof} \approx 0.9$. Given two values $x_{r_1}$ and $x_{r_2}$, the exponents can be easily computed as :

\begin{equation}
\label{eq:r1r2}
\begin{array}{r c c}
z \nu &=& \dfrac{x_{r_2} - x_{r_1}}{r_1 x_{r_1} - r_2 x_{r_2}} \equiv a, \\
\\
d \nu - 2 \beta &=& \dfrac{(r_2-r_1)x_{r_1} x_{r_2}}{r_2 x_{r_2} - r_1 x_{r_1}} \equiv b.
\end{array}
\end{equation}

\noindent
According to the above expressions, we obtain $a = 2.17(8)$ and $b = 1.75(5)$. Furthermore, with either the $r_1$- or the $r_2$-quench, one can use $f_1$ scaling with 2-parameter fitting to obtain $\beta/\nu$ and $z + 1/\nu$, as indicated by Figs.~\ref{fig02} and \ref{fig06}. We use the $f_1$ scaling from the $r=1$ quench, treating all exponents as unknown and performing a 2-parameter fitting for $p_1 = z+1/\nu$ and $p_2 = \beta/\nu$ and we obtain $p_1 = 3.16(5)$, $p_2 = 0.13(1)$ with $\chi^2/\text{dof} \approx 1.0$. Combining with the results from Eq.~(\ref{eq:r1r2}), one can then solve for $z$, $\nu$, and $\beta$:

\begin{equation}
\left.
\left \{
\begin{array}{r c c}
z+1/\nu &=& p_1, \\
z \nu &=& a, \\
\beta/\nu &=& p_2, 
\end{array}
\right.
\hspace{10pt}
\Rightarrow 
\hspace{10pt}
\right.
\left \{
\begin{array}{r c c}
z &=& 2.16(4), \\
\nu &=& 1.00(3),\\
\beta &=& 0.13(1).
\end{array}
\right.
\end{equation}
These exponents all agree with their known or expected (in the case of $z$) values within the error bars,
which were estimated by introducing Gaussian noises to the fit parameters $a$, $p_1$, and $p_2$ and solving the equations repeatedly.

This method should be particularly useful in cases where it is difficult to reach the adiabatic limit and carry out standard finite-size scaling techniques around $T_c$, e.g., for frustrated systems such as spin glasses.\cite{katzgraber06}

\subsection{Swendsen-Wang dynamics}
\label{subsec:sw}

Due to the rather large dynamic exponent, the Metropolis algorithm suffers significantly from critical slowing down. Physically, the slow dynamics originates from
the inability of single-spin (or any local)  updates to quickly change the structure of configurations with large clusters. In the SW algorithm,\cite{PhysRevLett.58.86} a spin configuration is decomposed into clusters using bond variables introduced through the Fortuin-Kasteleyn transformation.\cite{Physica.57.536, PhysRevE.68.056701} A broad range of cluster sizes appear according to Coniglio-Klein droplet theory,\cite{J.Phys.A.13.2775} and the algorithm is therefore much more efficient (has a much smaller dynamic exponent) than the Metropolis scheme.

In the SW algorithm, one unit of time is defined as decomposing the all spins in a configuration into clusters, using bonds set between same-oriented spins
with probability $P=1-{\rm e}^{-2J/T}$. Each spin uniquely belongs to one of the clusters (with spins having no connected bonds treated as clusters of size $1$) and each 
cluster is flipped independently with probability $1/2$. In the quench process we again start at $T_i = 1.5T_c$ and stop exactly at the known $T_c$, repeating 
the procedure thousands of times for averaging.

The dynamic scaling,  summarized as Eq.~(\ref{eq:2scaling_functions}), is independent of the underlying updating scheme, except for the value of $z$. We can therefore carry out the same kind of non-equilibrium quench process as in the previous subsection to study SW dynamics. Here we will focus on the linear quench ($r=1$) of the 2D Ising model.

\begin{figure}
   \includegraphics[width=8.4cm,clip=true]{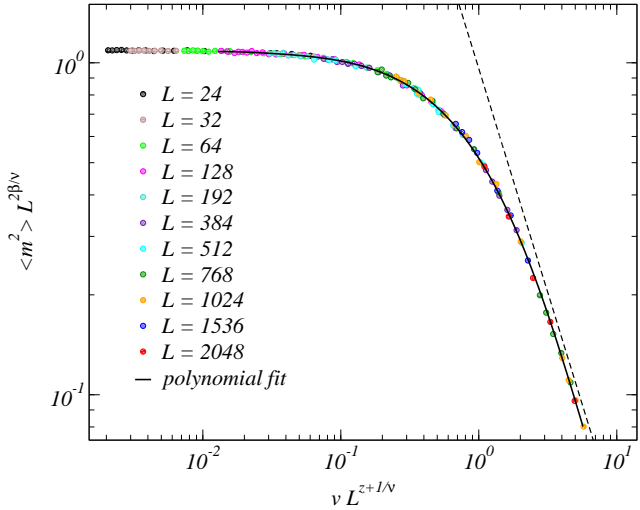}
   \caption{ (Color online) Results of linear quenches with SW dynamics of the 2D Ising model. The magnetization squared and the quench velocity are rescaled according to 
             the first line in Eq.~(\ref{eq:2scaling_functions}), resulting in data collapsing onto the scaling function $f_1$. A high-order polynomial
             was fitted to the data and the dynamic exponent was adjusted to optimize this fit, giving the optimal exponent $z_{\rm SW} = 0.297(3)$ with 
             $\chi^2/{\rm dof} \approx 1.0$. The dashed line indicates the predicted power-law behavior according to Eq.~(\ref{eq:power}) in the 
             universal scaling regime given the optimized value of $z_{\rm SW}$.}
  \label{fig08}
\end{figure}

\begin{figure}
   \includegraphics[width=8.4cm,clip=true]{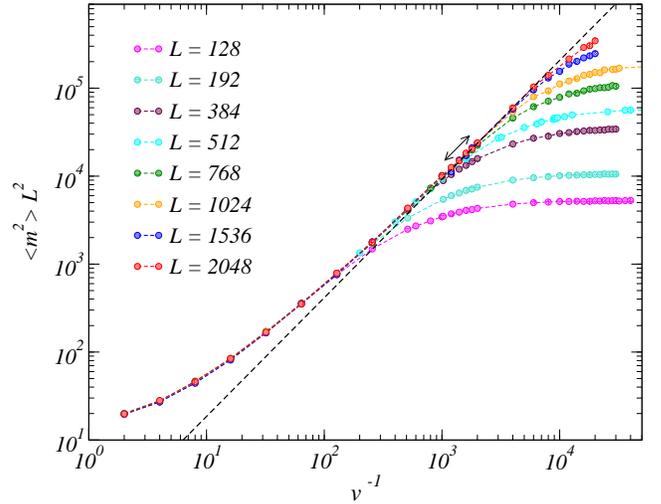}
   \caption{(Color online) Scaling collapse using the second line of Eq.~(\ref{eq:2scaling_functions}) to obtain the diabatic to power-law scaling function $f_2$ in the case of SW dynamics. The dashed line shows the slope expected with the dynamic exponent extracted in Fig.~\ref{fig08}. As in the case of Metropolis, one can also independently obtain $z$ by measuring the slope in the power-law regime. The line with arrow indicates the region in which linear fit is performed after taking log-log, we obtain $x = 1.35(4)$ with $\chi^2/\text{dof} \approx 0.9$, which implies $z_{\rm SW} = 0.29(4)$. } 
   \label{fig09}
\end{figure}

We again observe scaling collapse onto a scaling function $f_1$ according to the first line of Eq.~(\ref{eq:2scaling_functions}), as shown in Fig.~\ref{fig08}
(where we have not shown the diabatic data points, which deviate from the common scaling function---they will be analyzed further below). Here the dynamic exponent 
was again optimized to give the best fit. Due to the rather small value of the exponent in this case, $z_{\rm SW} \approx 0.3$, the universal power-law scaling 
regime is less accessible than in the Metropolis case. We therefore use a polynomial fit (to log-log data) in the whole region of the scaling variable in the
figure, instead of dividing it into two velocity regimes. Nonetheless, given the predicted power $x$, Eq.~(\ref{eq:power}), one can still test the consistency with the power-law behavior expected in the universal scaling regime after the optimized $z_{{SW}}$ has been obtained. The result for the dynamic exponent is $z_{\rm SW} = 0.297(3)$, which is consistent with Ref.~\onlinecite{PhysRevLett.62.163} (but with a smaller error bar). The dashed line in Fig.~\ref{fig08} shows the predicted power-law given the above value of the dynamic exponent. The agreement is indeed good for the right-most points. This behavior strongly supports the conventional critical dynamics with $z_{\rm SW} > 0$, instead of a logarithmic divergence of the time scale.\cite{J.Stat.Mech.2006.P05004} 

Note again that, for clarity, in Fig.~\ref{fig08} we have not shown the diabatic points deviating from the common scaling function. These data points are included in Fig.~\ref{fig09}, which shows a scaling collapse according to the second line of Eq.~(\ref{eq:2scaling_functions}). We can observe that the universal power-law regime is reached in a window of $1/v$ where the data is collapsed for the system sizes we have used. As demonstrated in Metropolis case, one can independently estimate $z_{\rm SW}$ by performing linear fit in the $f_2$ scaling after taking log-log, this procedure yields $x = 1.35(4)$ with $\chi^2/\text{dof} \approx 0.9$, which implies $z_{\rm SW} = 0.29(4)$. As in the Metropolis case, the $z_{\rm SW}$ extracted by $f_1$ and $f_2$ scalings are completely consistent.

\subsection{Wolff dynamics} 
\label{subsec:wolff}

The Wolff algorithm \cite{PhysRevLett.62.361} is an improvement of the SW algorithm. It is based on constructing single clusters according to the same bond rule as in
the SW algorithm, but each time starting from a random seed site (instead of one not previously visited when decomposing the whole system into non-overlapping
clusters in the SW algorithm) and flipping the clusters with probability one. The clusters are then on average larger than in the SW algorithm, and
the dynamic exponent is therefore normally smaller.\cite{J.Stat.Phys.58.1083} 

In order to compare the dynamics of the SW and Wolff algorithms it is important to treat the time-step in the latter in such a way that the number of spins
flipped is proportional to $N$. Clearly, above $T_c$ this is not the case for a single cluster, but one can still define the elementary unit of time as the
flipping of one cluster and subsequently rescale the time based on the average cluster size, so that an extensive number of spins are flipped in the rescaled
time unit. This is straight-forward in the equilibrium, where the scaling of the average cluster size is known in terms of critical exponents and the
Fourtuin-Kasteleyn mapping.\cite{Physica.57.536,J.Phys.A.13.2775} The critical Wolff cluster size scales as the magnetic susceptibility, $\chi \sim L^{\gamma/\nu}$. 
This implies that on average $\sim L^d/L^{\gamma/\nu}$ Wolff updates correspond to one MC step as defined in SW or Metropolis dynamics. Denoting by $z'_{W}$ 
the dynamic exponent measured using the single-cluster time unit and by $z_{W}$ the exponent corresponding to properly rescaled time, the relationship 
between these exponents is therefore \cite{J.Stat.Phys.58.1083}
\begin{equation}
\label{zshifting}
z_{W} = z'_{W} - (d-\gamma/\nu).
\end{equation} 
In non-equilibrium simulations the situation is more complicated, as we will see below. 

\begin{figure}
   \includegraphics[width=8.4cm, clip=true]{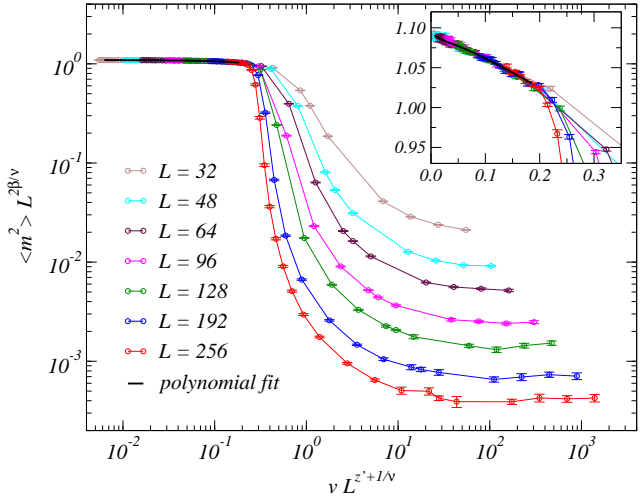}
   \caption{(Color online) Scaling collapse in the quasi-adiabatic regime (giving the scaling function $f_1$) for Wolff cluster dynamics with the time unit defined 
            as the flipping of a single cluster. Here the scaling collapse appears to break down at a singular point, as shown in greater detail in the inset. In the 
            regime where scaling collapse can be achieved, the optimized value of the dynamic exponent is $z'_{W} = 0.55(2)$ with $\chi^2/{\rm dof} \approx 1.0$.}
    \label{fig10}
\end{figure}

We here perform the same kind of linear quench with Wolff dynamics at different velocities as in the previous subsections for Metropolis 
and SW dynamics. We consider the elementary time unit as a single cluster flip and later discuss the subtleties involved in this definition. 

The scaling procedure of the squared magnetization expected to yield the function $f_1$ is shown in Fig.~\ref{fig10}. Here we observe a feature distinctively 
different from the Metropolis and SW cases: There is no universal scaling regime with power-law behavior. There is still a quasi-adiabatic regime where the data 
collapse well. We discuss this regime first and will return later to the lack of universal power-law scaling regime. 

In the quasi-adiabatic regime the rescaled squared magnetization is rather flat in Fig.~\ref{fig10}. Upon closer examination, as shown in the inset, there is still 
a clear drop when the scaled velocity approaches the region where the data-collapse breaks down. Interestingly, that break-down appears to take place at a single 
well defined point. Using a polynomial fit to the data before this point and optimizing the collapse by adjusting the dynamic exponent as in the previous cases,
we obtain $z'_{W} = 0.55(2)$. This again is the dynamic exponent measured according to the single-cluster definition of time, and to compare with Metropolis and 
SW dynamics the exponent should be shifted according to Eq.~(\ref{zshifting})---provided that the quench is sufficiently adiabatic throughout this regime.
Since for the 2D Ising model $\gamma = 7/4$ and $\nu=1$, we obtain $z_{W} = 0.30(2)$. This value is in good agreement with previous results, e.g., 
Ref.~\onlinecite{J.Stat.Phys.58.1083}, providing further confirmation of the quench process being effectively adiabatic in the regime where the 
scaling collapse occurs in Fig.~\ref{fig10}.

\begin{figure}
   \includegraphics[width=7.4cm, clip=true]{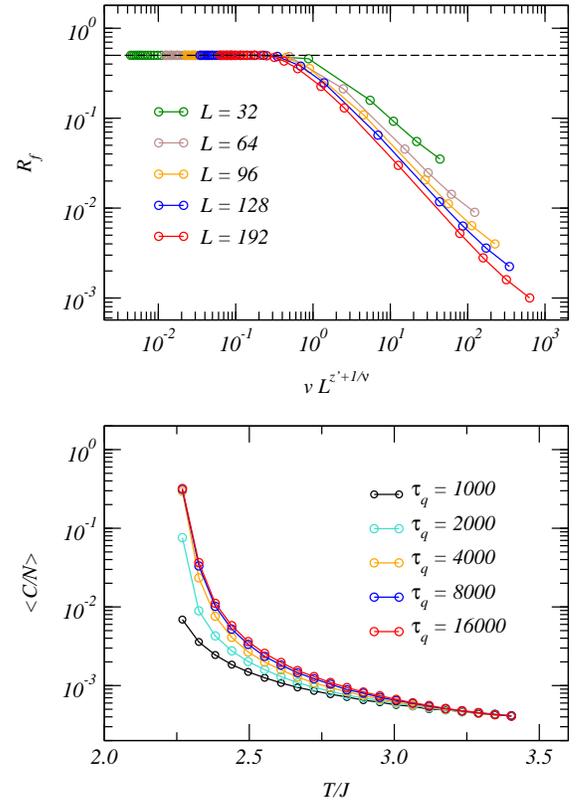}
   \caption{ (Color online) Bottom panel: Temperature dependence of the average cluster size $C$ relative to the system size $N$
             in Wolff-quenches at different velocities on a $128 \times 128$ lattice. 
             Top panel: The fraction of spins that are actually flipped with respect to the initial configuration after the entire quench process, graphed 
             as a function of the rescaled velocity with the dynamic exponent $z'_W=0.55$ as obtained in Fig.~\ref{fig10}. The fraction of flipped spins should 
             approach $1/2$ if the system at the initial and final times are completely decorrelated.}
    \label{fig11}
\end{figure}

Let us now discuss the break-down of scaling collapse and absence of a power-law regime at higher rescaled velocity. 
It seems clear that the break-down should be related to the single-cluster 
definition of the time unit. The typical size of the cluster is naturally associated with the temperature and the corresponding KZ correlation length, $\xi_{v}$,
reached at a given time step. This implies that at the early stage of the quench most of the system is left untouched by the Wolff construction, due to the 
small $\xi_{v}$ and cluster size. The growth of the cluster size versus $T$ as $T$ is decreased is of course slower than in the equilibrium. The bottom panel of 
Fig.~\ref{fig11} shows the average cluster size for different quench times as a fraction of the total system size for a system of size $128^2$. It is also 
illuminating to examine the fraction $R_f$ of spins actually flipped with respect to the initial configuration during the {\it entire} quench, i.e., counting the 
number of spins that are different in the initial and final configurations. If the simulation is ergodic within the total quench time, so that the initial and
final configurations can become completely decorrelated, this fraction should be very close to $1/2$ (strictly speaking, $R_f \to 1/2$ exponentially fast
for quench times much longer than the autocorrelation time). Furthermore, with any definition of the time unit where all spins are visited, the fraction 
approaches some $L$-independent constant, $R_f \in (0,1/2)$, when $v \to \infty$ (in practice, with our definitions, at $v=(T_i-T_c)/\tau_q$ with $\tau_q=1$;
the minimum quench time of one MC step). However, as shown in the top panel of Fig.~\ref{fig11}, with the single-cluster definition the flipped fraction 
decays sharply with increasing velocity and size. Interestingly, it reaches $1/2$ at a scaled velocity very close to the special point where the scaling 
collapse breaks down in Fig.~\ref{fig10}. It is clear that no quasi-adiabatic evolution, or even critical scaling, can take place if the scheme
effectively is non-ergodic, as the $R_f \to 0$ behavior indicates.

There is still of course a diabatic regime where in the high-velocity limit
the magnetization squared approaches its equilibrium value at the initial temperature. In this case, since the effect of the single-cluster flips in one unit 
of time changes with $T$, the velocity is not constant if one rescales to a time unit in which an extensive number of spins is flipped. Therefore, effectively, the 
procedure corresponds to a nonlinear quench protocol leading to an effectively much faster approach to the diabatic limit with increasing $v$ than in 
schemes based on usual definitions of the time unit with an extensive number of spin flips. With usual time definitions, for any system size $L$ one can
reach any configuration, in principle, in a single time steps, while with the Wolff algorithm, in the diabatic regime, the number of steps (flipped clusters) 
needed for ergodic sampling increases with the system size and also with the velocity (since the clusters increase in size with decreasing velocity).

\begin{figure}
   \includegraphics[width=8.4cm,clip=true]{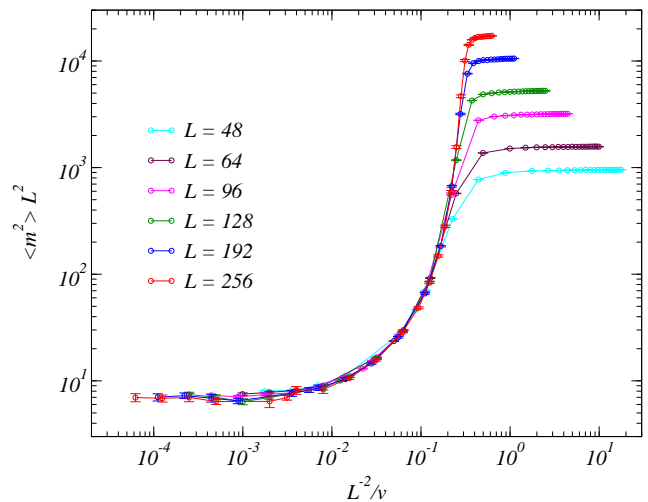}
   \caption{(Color online) Scaled squared magnetization in the diabatic regime, using a velocity rescaling of the form expected in this regime
            (accounting for the size of the Wolff-clusters decreasing with the system size as $L^{-2}$ for fixed $v$).}
   \label{fig12}
\end{figure}

Despite the peculiarities of the Wolff time unit, we can still attempt to rescale the data in Fig.~\ref{fig10} using the same diabatic approach as in the 
SW and Metropolis cases, to obtain the scaling function $f_2$ in Eq.~(\ref{eq:2scaling_functions}) for Wolff dynamics. However, in this case we have to modify the 
argument $a^{-(zr+1/\nu)}/v=v^a_{KZ}/v$ of the scaling function, because the effective velocity is normalized up by a factor, the inverse average fraction of 
flipped spins in a time step, which, as we have seen above, vanishes with increasing $L$ for fixed $v$. Since we are analyzing the diabatic regime, where the cluster 
size should be $L$-independent for sufficiently large $L$, the flipped fraction of spins in one step should scale as $L^{-d}$, and, thus, we should let $v \to vL^{d}$ 
in the scaling analysis. Setting the lattice scale parameter $a=1$ as before, we therefore expect scaling collapse with $f_2(L^{-d}/v)$, here with $d=2$. 
Indeed, as shown in Fig.~\ref{fig12}, graphing $\langle m^2\rangle L^2$ versus $L^{-2}/v$, the data collapse almost perfectly, down to a velocity where the scaling 
function appears to diverge in the thermodynamic limit (with the quasi-adiabatic plateaus splitting off later as $L$ grows). 

The above analysis shows that, even with the subtleties of the single-cluster time unit in the Wolff algorithm, there are still two well defined slow and fast 
regimes, where essentially perfect data collapse onto functions $f_1$ and $f_2$ can be achieved. Unlike the cases of Metropolis and SW dynamics, these scaling 
functions do not have any universal power-law parts connecting them in cross-overs, but instead they both break down in a singular manner with one type of scaling 
replaced by a completely different kind of scaling. In terms of rescaling of the time unit of the single-cluster Wolff steps, we have demonstrated that on the 
adiabatic side it is with the standard factor (same as in the equilibrium), $t \to tL^{-(d-\gamma/\nu)}$, while on the diabatic side it is just $t \to tL^{-d}$. 
The failures of these time rescalings at singular points leads us to conclude that Wolff dynamics is associated with a {\it dynamic phase transition}, and this 
transition is related to a sudden effective loss of ergodicity as a function of the velocity in fast quenches. 

\section{Summary and Discussion}
\label{sec:summary}

We have demonstrated a non-equilibrium quench approach and associated dynamic scaling scheme for studying the scale-invariant universal behavior 
and various cross-over behaviors when approaching critical points of classical phase transitions. 
Using three different variants of MC dynamics---Metropolis, SW, and Wolff---we
demonstrated that the order parameter (the squared magnetization) is governed by two different scaling functions describing  quasi-adiabatic
(including fully adiabatic) and diabatic (including extreme diabatic) evolution from an initial paramagnetic state to the critical point.
In all cases we have studied, the two scaling functions capture the dynamic behavior for the entire range of velocities
$v \in [0,\infty)$ for all system sizes (up to very small subleading finite-size corrections also present in the equilibrium). 
This complete characterization of the non-equilibrium scaling for several dynamic schedules was the main result of the paper. In addition, we showed that 
the quench scheme can also be used to extract accurate values of the dynamic exponent for given combinations of models and dynamics. In the main part of the paper
we used the standard 2D Ising model, but we have also investigated the 3D and fully-connected (infinite-dimensional) variants and report results 
for them in Appendix \ref{appb}. We summarize our results for the dynamic exponents in Table.~\ref{ztable}. 

\begin{table}
  \begin{tabular}{ |c|c|c| }
  \hline
  Dynamics & \hspace{1pt} Model \hspace{1pt} & \hspace{2pt} $z$ \hspace{2pt} \\ \hline
  Metropolis & 2D  &  $2.1767(5)$ \\ \hline
  \multirow{3}{*}{Swendsen-Wang} 
  & 2D  &  $ 0.297(3)$ \\ 
  & 3D  &  $ 0.53(1) $ \\
  & fc &  $ 1.2(2)$ \\ \hline
  \multirow{3}{*}{Wolff} 
    & 2D & $0.30(2)$ \\ 
    & 3D & $0.24(2)$ \\ 
    & fc & $0.04(4)$ \\
   \hline
\end{tabular}
\caption{Dynamic exponents obtained using either $f_1$ or $f_2$ scaling for Ising models in two and three dimensions, as well as the fully-connected
(fc) model (infinite-dimensional). The Metropolis dynamic exponent for 2D case quoted above is from $f_2$ scaling of $r=2$ quench, which yields the best estimate so far. The exponents for Wolff dynamics have been shifted using Eq.~(\ref{zshifting}) to account for the single-cluster
definition of the time unit of the simulations.}
\label{ztable}
\end{table}

In this paper we performed linear and non-linear quenches to exactly the critical point $T_c$ and observed excellent scaling in both cases. 
The quasi-adiabatic and diabatic scaling functions can be described perturbatively in $v$ and $1/v$, respectively, for small values of these
parameters. These regimes are normally (for SW and Metropolis dynamics) smoothly connected to each other via cross-overs to a universal, non-perturbative 
power-law scaling regime that can be described by either function. However, with Wolff dynamics, the two scaling regimes are separated in a singular 
manner and there is no power-law regime. This can be traced to the single-cluster definition of the time unit in the Wolff algorithm, which for a linear 
quench leads to an effectively non-linear, ultrafast approach to the diabatic limit, where the scheme becomes effectively non-ergodic. It
is remarkable that the loss of ergodicity takes place in such a singular way, and not through a smooth cross-over. The singular change in 
the scaling function can be interpreted as a dynamic phase transition. 

An issue with the non-linear quenches of the form (\ref{rquench}) is that the critical point has to be known exactly for the protocol to be asymptotically 
non-linear. If the critical point is not precisely known and the final point of the quench is therefore off the targeted critical value, the quench
ultimately becomes linear (if one stops below the true $T_c$) \cite{Dziarmaga2010} or doesn't reach $T_c$ at all (if one stops above the true $T_c$).
In this situation, assume that the final point of the non-linear quench is $T^\ast$, the offset from the critical point $|T_c - T^\ast| L^{1/\nu}$ enters as 
another argument of the scaling function (\ref{atclv}). As usual in the scaling theory, the shorter length scale dominates, and, provided $T^\ast$ is below 
$T_c$ and not too far off $T_c$, one should be able to observe non-linear scaling $r \not=1$ for some range of velocities before a cross-over to $r=1$ scaling. 
If $T^\ast > T_c$ there should instead be a cross-over into high-$T$ behavior with a finite correlation length. These cross-overs will be interesting 
targets for future studies.

Non-equilibrium relaxation from an ordered state has been widely used in the past to extract the dynamic exponents for ordered systems as well as 
spin glasses.\cite{Ozeki.J.Phys.A.40.R149, PhysRevE.56.2310, Parisi.J.Phys.A.30.7115} In our language, this corresponds to a sudden quench to
the critical point, $r=0$ in Eq.~(\ref{rquench}), starting from the ordered state (instead of starting from the disordered state, as we did in 
the present work). The ``velocity'' in this case is the inverse waiting time [unlike our definition (\ref{rquench}) where for $r \to 0$ there
is effectively waiting before a sudden quench and no waiting after], and the order parameter asymptotically decays as a power of the time. 
Normally the decay is studied for systems sufficiently large to effectively be in the thermodynamic limit for the time windows considered. We have not 
compared these approaches in terms of their abilities to extract high-precision values for the dynamic exponent, but at least naively it appears to us 
that it should be better to take advantage of finite-size and finite velocity scaling. In addition, for a linear quench one can also easily, without 
much additional computational effort, obtain results not only at a known (or approximately known) final critical point, but one can collect data also 
before the critical point is reached and continue past the critical point as well. This opens opportunities for other types of scaling studies in the 
vicinity of $T_c$, using Eq.~(\ref{atlv}) and its generalizations to incorporate both adiabatic and diabatic scaling functions.

Our value for the dynamic exponent for Metropolis updates of the 2D Ising model is in good agreement with the best known value.\cite{PhysRevB.62.1089} 
As we mentioned in Sec.~\ref{subsec:z}, for cluster dynamics the value of $z$ has been a matter of debate for some time. For SW dynamics, it was 
claimed in Ref.~\onlinecite{J.Stat.Mech.2006.P05004} that $z_{\rm SW} = 0$ for the 2D Ising model. However, based on our approach, a nonzero $z_{\rm SW}$ is clearly 
shown, not only in the scaling collapse of Fig.~\ref{fig08} but also as indicated by the consistent pow-law behavior in the universal scaling regime. 
For Wolff dynamics of the 2D Ising model, it was reported in Ref.~\onlinecite{J.Stat.Mech.2006.P05004} that $z_{W} = 1.19(2)$, which is significantly higher 
than the value obtained in Ref.~\onlinecite{J.Stat.Phys.58.1083}. The latter is consistent with our result in Table \ref{ztable}. 
As pointed out in Ref.~\onlinecite{J.Stat.Mech.2006.P05004}, $z_{W}$ computed with standard relaxation from an ordered state may in practice 
be sensitive to the initial state, unless extremely long times are considered. Furthermore, the result may also depend on the targeted 
observable.\cite{Ossola.Nucl.Phys.B.691.259} In this sense we think our approach is more stable in practice and has useful features for 
self-consistency checks, e.g., the same power laws appearing in all three dynamical regimes in Eq.~(\ref{eq:velocity_regimes}).

We have demonstrated that the dynamic exponent in principle can be extracted by two different kinds of scaling collapses, especially when the static exponents are already known, given either the quasi-adiabatic function $f_1$ or the diabatic function $f_2$ in Eq.~(\ref{eq:2scaling_functions}). Throughout the demonstration we show that the results of $z$ obtained by $f_1$ and $f_2$ are completely consistent. Since the diabatic quenches are very fast in comparison to the quasi-adiabatic ones, it is more tempting to focus on the universal power-law scaling before the cross-over into the quasi-adiabatic behavior. Apart from the savings in raw computer resources, the data-collapsing 
procedure for $f_2$ in the universal scaling regime requires no knowledge or optimization of critical exponents; one simply plots $\langle m^2 \rangle N$ versus $v^{-1}$, 
in the style of Fig.~\ref{fig04}, Fig.~\ref{fig07}, or Fig.~\ref{fig09}, and uses linear fit to extract the slope $x$ of the collapsed data on the log-log plot. The
resulting $x$ of course still is a combination of the critical exponents (\ref{eq:power}) and one needs some further steps to disentangle them.

It is very interesting to note that all the static exponents can be extracted along with $z$ by combining results from two different quench protocols characterized by two different values of $r$ in Eq.~(\ref{rquench}), as we demonstrated in Sec.~\ref{sec:combination}. This may potentially be very beneficial to systems such as spin glasses, where the large dynamic exponent makes it very difficult to carry out equilibrium calculations for large systems.\cite{katzgraber06} In our proposed method above, the need to equilibrate configurations at and close to the glass transition is completely circumvented, as one can start from some elevated temperature, where the equilibration is fast, and any slowing down ``problem'' just reflects the underlying dynamic exponent and manifests itself in the form of the desired exponent $x$. 

We also point out that the non-equilibrium scheme discussed here is not restricted only to classical phase transitions, but also applies to quantum
phase transitions, which can be studied, e.g., with the quantum MC simulation schemes recently developed in Refs.~\onlinecite{PhysRevB.84.224303,qaqmc} 
for evolution in imaginary time. Some results for transverse-field Ising models were already reported in the above papers.

In this paper, we have discussed temperature quenches, but the same framework is also applicable to quenching, e.g., an external field.
Beyond the critical scaling discussed in this paper, such quenches allow one to investigate numerous aspects of first-order phase transitions and 
hysteresis in the Ising model, as was already done in some cases.\cite{PhysRevE.77.030103R, PhysRevB.70.132403, PhysRevE.52.1399, Gong10}

\begin{acknowledgments}
We would like to thank Bill Klein and Fan Zhong for stimulating discussions and Wenan Guo for comments on the manuscript. 
This work was supported by the NSF under Grant No.~PHY-1211284.
\end{acknowledgments}


\appendix

\section{Derivation of the Kibble-Zurek velocity for non-linear quenches and construction 
of optimal adiabatic quench protocols}
\label{appa}

The original derivation of the KZ mechanism is not entirely satisfying, as it treats proportionality in a 
rather cavalier (though ultimately correct) way. For instance, in the first argument, Eq. (\ref{eq:quench_time}), 
for the total time required for a linear quench to stay in equilibrium, there is an apparent dependence on 
the initial temperature $T_i$, which disappears in the subsequent analysis. Furthermore, if the initial temperature
is very close to the final one, it appears that one should take into account that the system is already from
the outset almost in equilibrium at the final temperature, and, hence, the time required to stay in equilibrium should
be shorter than suggested by Eq.~(\ref{eq:quench_time}). More generally, any quench taking place over a finite time 
can be seen as a series of small sudden quenches, and in each of them the time to equilibrate should be much smaller
than Eq. (\ref{eq:quench_time}). From the outset it is not clear whether the total, integrated time of a quench 
staying in equilibrium is really the same as Eq. (\ref{eq:quench_time}) and its nonlinear generalizations. We will
here show very explicitly that this is true for power-law quenches. The calculations can easily be generalized for
any protocol.

We also note that, for nonlinear quenches of the form (\ref{rquench}), when $r\to 0$ the protocol effectively turns 
into waiting for the full time $\tau_q$ and then quenching suddenly to $T_c$, while $r \to \infty$ corresponds to the 
reversed situation; first quenching suddenly to $T_c$ and then waiting a time $\tau_q$. It is then interesting to see how $r$ 
enters in the proportionalities, e.g., in the finite-size KZ velocity Eq. (\ref{eq:quench_velocity}). Our derivations below
gives several prefactors of the proportionalities, which, among other things, can be useful for constructing optimized
quench protocols.

\subsection{Kibble-Zurek velocity}

\begin{figure}
   \includegraphics[width=6.5cm, clip=true]{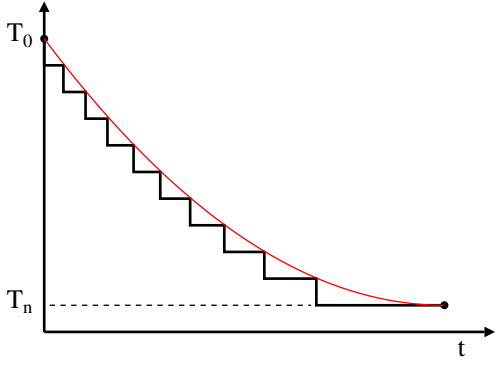}
   \caption{(Color online) Discretized quench protocol in which the continuous time dependence of the temperature is replaced by
            a series of sudden quenches in which the temperature is changed by a constant amount $\Delta_T$.} 
    \label{fig13}
\end{figure}

We here treat a power-law quench of the form (\ref{rquench}) for arbitrary $r$ (including $r < 0$) and any final time $\ge T_c$. We imagine 
dividing the temperature window of interest into $n$ intervals $\Delta_T$ between points $T_0,T_1,\ldots,T_n$, with $T_0$ the initial 
temperature (at which the system is presumed to be in thermal equilibrium before the quench starts) and the final temperature $T_n$, 
with $T_c \le T_n < T_0$. We replace the smooth quench by a series of $n$ steps consisting of the temperature dropping suddenly by 
the fixed amount $\Delta_T$ from $T=T_i$ to $T_{i+1}$, followed by a waiting time $\Delta_t(T)$ determined by the protocol and
$\Delta_T$. This is illustrated in Fig.~\ref{fig13}. We will eventually take the limit $\Delta_T \to 0$ to recover the 
smooth quench process.

With the protocol (\ref{rquench}) generalized to final time $T_n$, the time versus temperature is given by
\begin{equation}
\label{appa:t1}
t = \tau_q - v^{-1/r}(T-T_n)^{1/r},
\end{equation}
and therefore the time taken for the $i$:th quench step, where the temperature decreases from $T_i$ to $T_{i+1}$ by the
small amount $\Delta_T$, is given by
\begin{equation}
\Delta_t(T_i) = \frac{\Delta_T}{rv^{1/r}} (T_i-T_n)^{1/r-1}. \label{appa:dt}
\end{equation}
Now consider the relaxation time. Since the steps are small, if the system was in equilibrium at $T_i$ it is almost
in equilibrium after the temperature change, and the time $\Delta_\tau(T_i)$ to equilibrate at the new temperature should 
be given by the time taken to increase the correlation length from $\xi(T_i)$ to $\xi(T_{i+1})$, which should be just the 
differences in the equilibration times at the two temperatures;
\begin{eqnarray}
\Delta_\tau(T_i) & = & \xi_T^z(T_{i+1}) - \xi_T^z(T_{i}) \nonumber \\
                 & = & (T_{i+1}-T_c)^{-\nu z} - (T_{i}-T_c)^{-\nu z}\nonumber \\
                 & = & -\Delta_T \frac{d}{dT} (T-T_c)^{-\nu z} |_{T=T_i} \nonumber \\
                 & = & \Delta_T \nu z  (T_i-T_c)^{-\nu z -1}. \label{appa:dtau}
\end{eqnarray}
The criterion for staying in equilibrium is now simply obtained by comparing this incremental relaxation time with the
time taken for the $i$:th step according to Eq.~(\ref{appa:dt}). We can use this kind of comparison in different ways, e.g.,
to find the velocity needed to stay in equilibrium all the way to the final temperature $T_n$ or to find the temperature
$T_i$ at which the system falls out of equilibrium. Here we explicitly consider quenches all the way to the critical temperature;
$T_n=T_c$. 

We can first ask what the requirement is for staying in equilibrium during the whole process. 
The criterion for this is that $\Delta_t(T_i) > \Delta_\tau(T_i)$ for all $i$, which, according to 
Eqs.~(\ref{appa:dt}) and (\ref{appa:dtau}) means that 
\begin{equation}
\frac{1}{r v^{1/r}} (T-T_n)^{1/r-1} > \nu z (T-T_c)^{-\nu z -1},
\label{appa:crit}
\end{equation}
where we have implicitly taken the limit $\Delta_T \to 0$. Here we can disregard the prefactors and just conclude that
the quench can stay adiabatic all the way down to $T_c$, provided that 
\begin{equation}
\frac{1}{r} < -\nu z,
\label{appa:adicrit}
\end{equation}
i.e., $r$ has to be negative and it takes an infinitely long time to actually reach $T_c$---in this case the protocol formally has
to be modified, with the total time (\ref{appa:t1}) spent to get down to temperature $T$ becoming
\begin{equation}
\label{appa:t2}
t = \frac{v^{1/|r|}}{(T-T_c)^{1/|r|}},~~~(r < 0).
\end{equation}
This kind of ultra-slow quench was recently considered in work by Chandran {\it et al},\cite{PhysRevB.86.064304} who also obtained,
through a different derivation, the adiabaticity criterion (\ref{appa:adicrit}).

When the quench parameter $r$ is larger than the above threshold value, the system will fall out of equilibrium at some temperature, when
$\Delta_\tau(T) > \Delta_t(T)$. To analyze this condition we keep all the prefactors in Eqs.~(\ref{appa:dt}) and (\ref{appa:dtau}).
The threshold criterion for staying in equilibrium according to Eq.~(\ref{appa:crit}) for given $r$ can then be written as 
(with $T_n=T_c$ and taking the $\Delta_T \to 0$ limit)
\begin{equation}
\nu z  (T-T_c)^{-\nu z -1} \sim
\frac{1}{rv^{1/r}} (T-T_c)^{1/r-1},
\end{equation}
which we simplify as
\begin{equation}
(T-T_c)^{\nu z +1/r} \sim \nu z r v^{1/r}.
\end{equation}
From this proportionality expression we can extract the temperature at which the system falls out of equilibrium for fixed $r$ and $v$. 
We can also extract the critical velocity required to stay in equilibrium down to temperature $T>T_c$, i.e., the KZ velocity,
\begin{equation}
v_{\rm KZ}(T) \sim (\nu z r)^{-r} (T-T_c)^{\nu z r + 1},
\label{kz_rel}
\end{equation}
which agrees with the $r=1$ result stated in Eq.~(\ref{eq:quench_velocity}). From this result we can also obtain an $L$-dependent 
KZ velocity (threshold for staying in equilibrium all the way down to $T_c$) by replacing $(T-T_c)^{-\nu}$ by $L$, giving
\begin{equation}
v_{\rm KZ}(L) \sim (\nu z r)^{-r} L^{-(z r + 1/\nu)},
\end{equation}
in agreement with Eq.~(\ref{eq:vkz}). However, we also now see that there is a prefactor
which becomes very small for large $r$, especially if $\nu z$ is large (as is the case, e.g., for spin glasses). Of course
the power of $T-T_c$ or $L$ is still dominant, but in some cases, e.g., for optimization purposes, it can be important 
to consider the prefactor as well. Similar $r$-dependent prefactors have been derived in a different way in the context of 
optimized passages through quantum-critical points ($T=0$).\cite{barankov08}

\subsection{Optimal protocol}

Eq.~(\ref{kz_rel}) gives the relation between the correlation length at the point where the adiabatic approximation breaks down 
and the generalized velocity of the quench:
\begin{equation}
\xi_{v}\sim{1\over |T-T_c|^\nu}\sim (\nu z r)^{-r/(z r+1/\nu)} v^{-1/(z r + 1/\nu)}.
\end{equation}
It is interesting that this simple expression can be immediately used to find the optimal adiabatic protocol maximizing the correlation 
length for a fixed quench time $\tau_q$. Expressing $v\sim 1/\tau_q^r$ we find that
\begin{equation}
\xi_{v}(\tau) \sim \left(\tau_q\over \nu z r\right)^{-r/(z r+1/\nu)}.
\end{equation}
This expression coincides with the one obtained in Ref.~\onlinecite{barankov08} for quantum quenches. The optimal value of the exponent $r$ 
can now be obtained by extremizing $\xi_{v}$ with respect to $r$ at fixed duration $\tau_q$, which is equivalent to the condition
\begin{equation}
{\nu  r\over (\nu z r +1)^2}\log\left({\tau_q\over \nu z r}\right)=1.
\end{equation}
In the long time limit ($\tau_q\gg 1$) this equation can be approximately solved, giving
\begin{equation}
r\sim {1\over \nu z} \log (\tau_q),
\label{appa:optr}
\end{equation}
which is exactly the optimum power obtained in Ref.~\onlinecite{barankov08} for quantum systems, and which was also subsequently obtained numerically 
for classical systems with Langevin dynamics.\cite{power13} Here we have obtained this classical result from the very simple but rigorous arguments of
successive equilibration illustrated in Fig.~\ref{fig13}. 

Note that in the limit $\tau_q\to \infty$ in Eq.~(\ref{appa:optr}), the optimal parameter $r\to \infty$, which, as discussed in the beginning of 
this Appendix, turns the protocol into a sudden quench followed by a waiting time $\tau_q$. However, since the divergence of $r$ is only logarithmic in
$\tau_q$, and $\nu z$ can be large, in practice, given a finite time resource $\tau_q$, the optimal protocol can still be very far from the
sudden quench.

\section{Higher-dimensional models}
\label{appb}

\begin{figure}
   \includegraphics[width=8.4cm, clip=true]{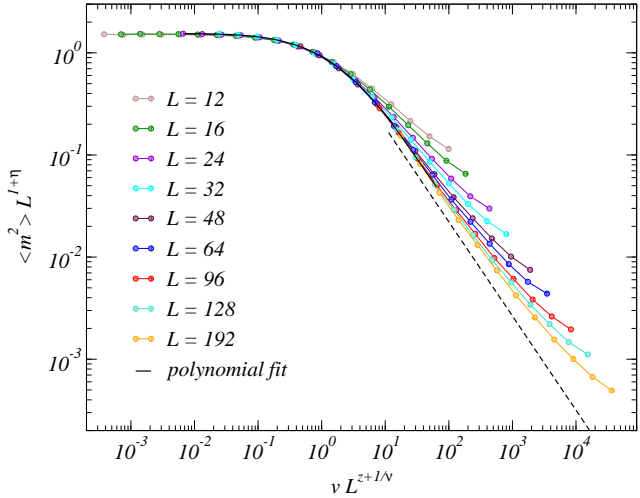}
   \caption{(Color online) Optimized scaling collapse for linear SW quenches of the 3D Ising model, giving $z_{_{SW}}=0.53(1)$ with $\chi^2/{\rm dof} \approx 1.1$. 
           The dashed line shows the anticipated power-law asymptotic behavior in the universal scaling regime, with the above value of the dynamic exponent
           and the slope $x$ given by Eq.~(\ref{eq:power}).} 
    \label{fig14}
\end{figure}

Despite the similar dynamic exponents of the SW and Wolff algorithms when applied to the 2D Ising model, $z_{{SW}} \approx z_W \approx 0.3$, the
degree of critical slowing-down with these algorithms can be very different in higher dimensions. 
To demonstrate that the scaling schemes developed and tested in the bulk of the paper 
also apply beyond the simple 2D Ising model, and to further examine the peculiarities of the Wolff algorithm discovered in Sec.~\ref{subsec:wolff}, 
we here present results of linear SW and Wolff quenches of Ising models in higher dimensions. The resulting dynamic exponents 
are listed in Table \ref{ztable}.

For the 3D Ising model, with the Hamiltonian (\ref{eq:Hising}) defined with nearest-neighbor interactions on the simple cubic lattice, 
numerical estimates for the critical point $T_c$ and the exponents are known to rather high precision;\cite{J.Stat.Phys.42.823, PhysRevB.59.11471} 
$J/T_c = 0.22169(2)$, $\nu = 0.6298(5)$, and $\eta = 0.0366(8)$. Given these exponents, we use the exponent relation $2\beta/\nu=1+\eta$ in the $r=1$ 
dynamic scaling relation (\ref{eq:m2scaling}). 

We write the Hamiltonian for the fully-connected (or infinite-dimensional) Ising model as
\begin{equation}
\label{fcising}
H = - \frac{J}{N-1} \sum_{i=1}^N \sum_{j>i} \sigma_i \sigma_j,
\end{equation}
where the coupling is normalized by the system size. Since mean-field theory becomes exact for this model when $N \to \infty$, we have $T_c/J= 1$, 
and the critical exponents are $\nu = 1/2$, $\beta = 1/2$. To apply scaling forms such as Eq.~(\ref{eq:2scaling_functions}) in this case we 
have to use the upper critical dimension of the Ising model, $d_c=4$, to define the effective system length as $L_{\rm eff} = N^{1/d_{c}}$ and use $d=d_c$.

\begin{figure}
   \includegraphics[width=8.4cm, clip=true]{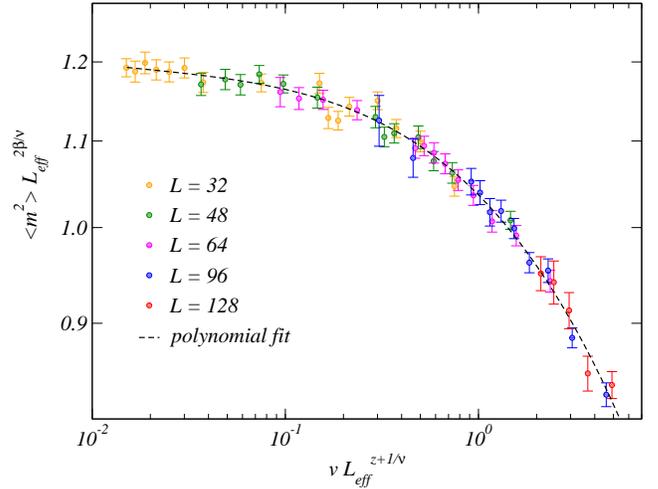}
   \caption{(Color online) Optimized log-log scaling collapse for linear SW quenches of the fully-connected Ising model, 
          giving $z_{_{SW}}=1.2(2)$ with $\chi^2/{\rm dof} \approx 0.8$.}           
    \label{fig15}
\end{figure}

\subsection{Swendsen-Wang dynamics}
\label{appendix_SW}

Data collapse for SW dynamics on the 3D model is shown in Fig.~\ref{fig14}. Fitting a polynomial to the quasi-adiabatic region and
adjusting the exponent to optimize the collapse as before gives $z_{_{SW}} = 0.53(1)$, this is in good agreement with Ref.~\onlinecite{PhysRevLett.68.962}.
The collapsed region where this fitting procedure was carried out
corresponds mainly to the quasi-adiabatic regime, but a cross-over to a power-law regime, with the slope consistent with the expected exponent, is also 
clear for the larger system sizes.

The same kind of scaling collapse for the fully-connected Ising model is shown in Fig.~\ref{fig15}. Here we have much less data, 
but, focusing on the quasi-adiabatic regime, we can observe scaling collapse with a dynamic exponent $z_{_{SW}} = 1.2(2)$. 
This in good agreement with mean-field calculation,\cite{J.Stat.Phys.58.1083} according to which $z_{{MF}}=1$.

\subsection{Wolff dynamics}
\label{appendix_Wolff}

\begin{figure}
   \includegraphics[width=8.4cm, clip=true]{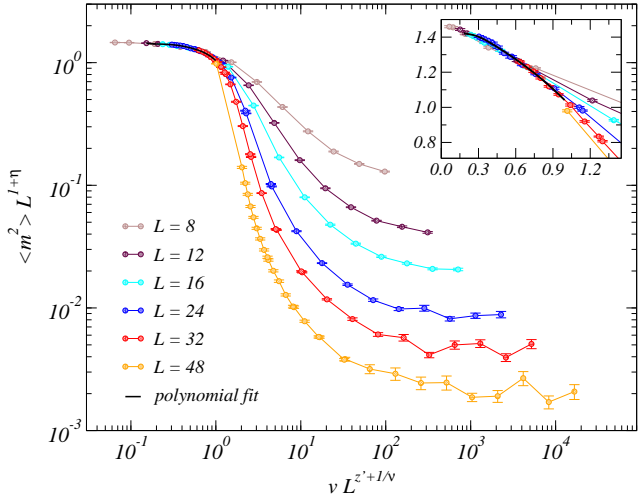}
   \caption{(Color online) Optimized scaling collapse in the quasi-adiabatic regime of the 3D Ising model with Wolff dynamics. The optimal dynamic exponent
            with the single-cluster time unit is $z'_{_W} = 1.27(2)$ with $\chi^2/{\rm dof} \approx 1.0$. The shifted value according to Eq.~(\ref{zshifting}) is
            $z_{_W} = 0.24(2)$. The inset shows more details of the data (but on a lin-lin plot) around the point where the data collapse breaks down.}
    \label{fig16}
\end{figure}

As we saw in the Sec.~\ref{subsec:wolff}, Wolff dynamics on the 2D Ising model exhibits scaling collapse in the quasi-adiabatic and diabatic regimes, but the
smooth cross-over with power-law scaling in Eq.~(\ref{eq:velocity_regimes}) is lacking. It is interesting to investigate how this behavior evolves as the dimensionality 
is increased, which we do here by studying the 3D and fully connected models.

The analysis of the data in the quasi-adiabatic regime is presented in Fig.~\ref{fig16}. Here, again, the data collapse appears to break down essentially at a point,
with no apparent signs of any emergent power-law scaling behavior (although the point at which the curves split off from the scaling function appears to show
somewhat more finite-size drift than in Fig.~\ref{fig10}, where almost no drift can be seen). The dynamic exponent is $z'_{_W} = 1.27(2)$. 
To compare this with the exponent in SW dynamics, one again has has to shift the value according to Eq.~(\ref{zshifting}), which gives $z_{_W} = 0.24(2)$. 
This is close to the result obtained in Ref.~\onlinecite{Phys.Lett.B.228.379}. 

\begin{figure}
   \includegraphics[width=8.4cm, clip=true]{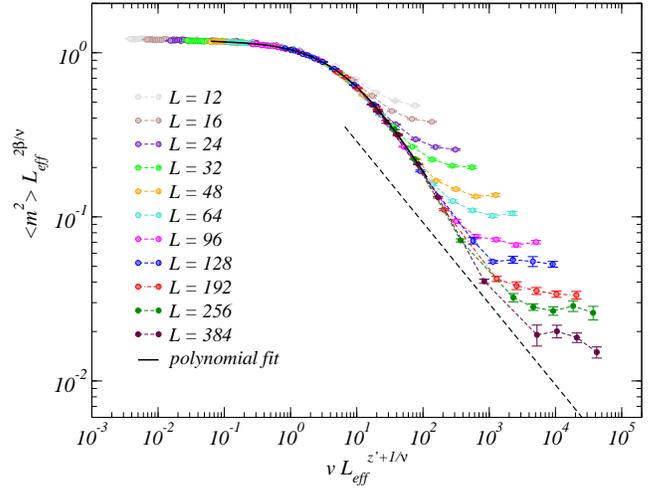}
    \caption{(Color online) Optimized scaling collapse in the quasi-adiabatic regime of the fully connected Ising model subject to Wolff 
             dynamics. The optimal dynamic exponent is $z'_{_W} = 2.04(4)$, or $z_{_W} = 0.04(4)$, with $\chi^2/{\rm dof} \approx 0.9$. The dashed
             line shows the expected behavior with $z'=2$ and mean-field static exponents ($\nu=\beta=1/2$) if the exponent relation (\ref{eq:power}) 
             is valid (which does not appear to be the case).} 
    \label{fig17}
\end{figure}

\begin{figure}
   \includegraphics[width=7.8cm, clip=true]{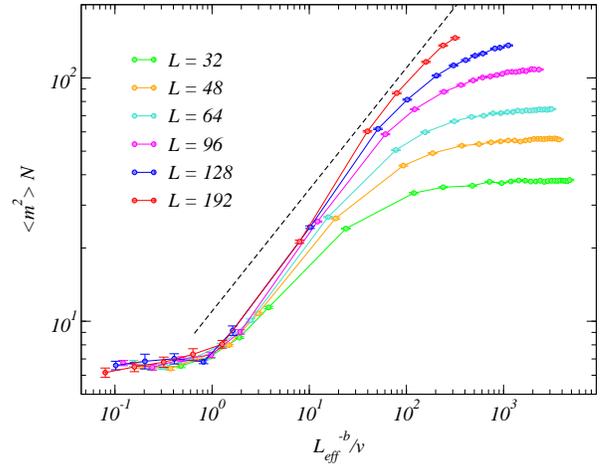}
    \caption{(Color online) The same data as in Fig.~\ref{fig17}, analyzed using diabatic scaling. Here the exponent $b \approx 1.2$ accounts for the
             growth of the Wolff-cluster size with $L_{\rm eff}$. It was optimized for the best data collapse. The dashed line shows the expected behavior
             with the exponent (\ref{eq:power}) with $z'=2$, that does not appear to apply here.}
    \label{fig18}
\end{figure}

Turning now to the fully connected Ising model, Fig.~\ref{fig17} shows the outcome of an optimized data collapse yielding $z'_{_W} = 2.04(4)$, or, after shifting 
the value according to Eq.~(\ref{zshifting}), $z_{_W} = 0.04(4)$. This confirms the expectation that the Wolff algorithm should be completely free from critical
slowing down in this case.\cite{J.Stat.Phys.58.1083} The figure also shows an interesting feature different from any of the other cases we have considered:
While the data collapse now does also extend to (apparently) arbitrarily high scaled velocities and the behavior does look like a power law, the slope
on the log-log plot is not what would be expected based on the relationship (\ref{eq:power}) with the dynamic exponent extracted in the optimization of
the data collapse (i.e., with $z'=2$ in place of $z$ in the expression for the exponent $x$). While we do not know the exact reason for this, one can 
suspect that it has to do with the non-locality of the model invalidating the arguments leading to the exponent relation (\ref{eq:power}), perhaps similar 
to violation of hyperscaling relations above the upper critical dimension.

In Sec.~\ref{subsec:wolff}, when analyzing the the diabatic regime for the 2D model, we had to rescale the velocity by a factor $L^d$ to account for the fact that the 
Wolff clusters stay constant in size for fixed $v$ when the system grows. In the fully-connected model, however, since the number of interacting bonds per site increases 
as $N$, the Wolff clusters should be expected to grow as well, as some power of the size. We have not investigated this behavior explicitly and therefore just assume
that it is power law and graph $\langle m^2\rangle L_{\rm eff}^{d_c}$ versus $L_{\rm eff}^{-b}/v$ ($=N^{-b/4}/v$), where $b$ is optimized and should be expected to be less 
than $d_u=4$ (since the clusters cannot grow faster than $N$). As shown in Fig.~\ref{fig18}, we can indeed achieve data collapse this way, with $b \approx 1.2$, 
although the subleading finite size corrections are very strong. This value of $b$ indicates that the Wolff clusters grow approximately as $\sim N^{0.3}$. 

Beyond the clearly diabatic behavior in Fig.~\ref{fig18} we cannot yet, for the range of system sizes considered, observe a clear power-law scaling regime, 
although the convergence of the data is certainty consistent with a power law. Again, as in Fig.~\ref{fig17} the behavior does not appear to be consistent 
with the expected exponent given by Eq.~(\ref{eq:power}), shown with the dashed line in Fig.~\ref{fig18}.




\begin{thebibliography}{99}
\bibitem{v_comment}
In Ref.\onlinecite{PhysRevB.81.012303} the protocol was defined as $T-T_c = v(\tau_q-t)^r/r!$. For simplicity we here leave out the factor $1/r!$.
\end{thebibliography}
\end{document}